\documentclass[longauth]{aa}  

\usepackage{graphicx}
\usepackage{hyperref}
\usepackage{euclid}
\usepackage{bm}

\usepackage{comment}
\usepackage[normalem]{ulem}
\usepackage[dvipsnames]{xcolor}
  
\usepackage[T1]{fontenc} 
\usepackage{braket}

\definecolor{lcol}{HTML}{006699} 

\usepackage[nameinlink,noabbrev]{cleveref}
\Crefname{equation}{Eq.}{Eqs.}
\Crefname{section}{Sect.}{Sects.}
\Crefname{figure}{Fig.}{Figs.}

\defcitealias{IST:paper1}{EC19}

\begin{document}

\title{\Euclid: impact of nonlinear prescriptions on cosmological parameter estimation from weak lensing cosmic shear\thanks{This paper is published on behalf of the Euclid Consortium.}}

\author{M.~Martinelli$^{1}$\thanks{\email{matteo.martinelli@uam.es}}, I.~Tutusaus$^{2,3}$, M.~Archidiacono$^{4,5}$, S.~Camera$^{6,7,8}$, V.F.~Cardone$^{9}$, S.~Clesse$^{10,11}$, S.~Casas$^{12}$, L.~Casarini$^{13,14}$, D. F. ~Mota$^{14}$, H.~Hoekstra$^{15}$, C.~Carbone$^{16}$, S.~Ili\'c$^{17,18,19}$, T.D.~Kitching$^{20}$, V.~Pettorino$^{12}$, A.~Pourtsidou$^{21}$, Z.~Sakr$^{19,22}$, D.~Sapone$^{23}$, N.~Auricchio$^{24}$, A.~Balestra$^{25}$, A.~Boucaud$^{26}$, E.~Branchini$^{9,27,28}$, M.~Brescia$^{29}$, V.~Capobianco$^{8}$, J.~Carretero$^{30}$, M.~Castellano$^{9}$, S.~Cavuoti$^{29,31,32}$, A.~Cimatti$^{33,34}$, R.~Cledassou$^{35}$, G.~Congedo$^{36}$, C.~Conselice$^{37}$, L.~Conversi$^{38,39}$, L.~Corcione$^{8}$, A.~Costille$^{40}$, M.~Douspis$^{41}$, F.~Dubath$^{42}$, S.~Dusini$^{43}$, G.~Fabbian$^{44}$, P.~Fosalba$^{2,3}$, M.~Frailis$^{45}$, E.~Franceschi$^{24}$, B.~Gillis$^{36}$, C.~Giocoli$^{34,46,47}$, F.~Grupp$^{48,49}$, L.~Guzzo$^{4,5,50}$, W.~Holmes$^{51}$, F.~Hormuth$^{52}$, K.~Jahnke$^{53}$, S.~Kermiche$^{54}$, A.~Kiessling$^{51}$, M.~Kilbinger$^{12,55}$, M.~Kunz$^{56}$, H.~Kurki-Suonio$^{57}$, S.~Ligori$^{8}$, P.B.~Lilje$^{14}$, I.~Lloro$^{58}$, E.~Maiorano$^{24}$, O.~Marggraf$^{59}$, K.~Markovic$^{51}$, R.~Massey$^{60}$, M.~Meneghetti$^{46,47}$, G.~Meylan$^{61}$, B.~Morin$^{62}$, L.~Moscardini$^{24,34,63}$, S.~Niemi$^{20}$, C.~Padilla$^{30}$, S.~Paltani$^{42}$, F.~Pasian$^{45}$, K.~Pedersen$^{64}$, S.~Pires$^{12}$, G.~Polenta$^{65}$, M.~Poncet$^{35}$, L.~Popa$^{66}$, F.~Raison$^{49}$, J.~Rhodes$^{51}$, M.~Roncarelli$^{24,34}$, E.~Rossetti$^{34}$, R.~Saglia$^{48,49}$, P.~Schneider$^{59}$, A.~Secroun$^{54}$, S.~Serrano$^{2,3}$, C.~Sirignano$^{43,67}$, G.~Sirri$^{68}$, J.-L.~Starck$^{12}$, F.~Sureau$^{12}$, A.N.~Taylor$^{36}$, I.~Tereno$^{69,70}$, R.~Toledo-Moreo$^{71}$, E.A.~Valentijn$^{72}$, L.~Valenziano$^{24,68}$, T.~Vassallo$^{48}$, Y.~Wang$^{73}$, N.~Welikala$^{36}$, A.~Zacchei$^{45}$, J.~Zoubian$^{54}$}

\institute{$^{1}$ Instituto de F\'isica T\'eorica UAM-CSIC, Campus de Cantoblanco, E-28049 Madrid, Spain\\
$^{2}$ Institute of Space Sciences (ICE, CSIC), Campus UAB, Carrer de Can Magrans, s/n, 08193 Barcelona, Spain\\
$^{3}$ Institut d’Estudis Espacials de Catalunya (IEEC), Carrer Gran Capit\'a 2-4, 08034 Barcelona, Spain\\
$^{4}$ Dipartimento di Fisica "Aldo Pontremoli", Universit\'a degli Studi di Milano, Via Celoria 16, I-20133 Milano, Italy\\
$^{5}$ INFN-Sezione di Milano, Via Celoria 16, I-20133 Milano, Italy\\
$^{6}$ INFN-Sezione di Torino, Via P. Giuria 1, I-10125 Torino, Italy\\
$^{7}$ Dipartimento di Fisica, Universit\'a degli Studi di Torino, Via P. Giuria 1, I-10125 Torino, Italy\\
$^{8}$ INAF-Osservatorio Astrofisico di Torino, Via Osservatorio 20, I-10025 Pino Torinese (TO), Italy\\
$^{9}$ INAF-Osservatorio Astronomico di Roma, Via Frascati 33, I-00078 Monteporzio Catone, Italy\\
$^{10}$ Namur Institute of Complex Systems (naXys), University of Namur, Rempart de la Vierge 8, 5000 Namur, Belgium\\
$^{11}$ Cosmology, Universe and Relativity at Louvain (CURL), Institut de Recherche en Mathematique et Physique (IRMP), Louvain University, 2 Chemin du Cyclotron, 1348 Louvain-la-Neuve, Belgium\\
$^{12}$ AIM, CEA, CNRS, Universit\'{e} Paris-Saclay, Universit\'{e} Paris Diderot, Sorbonne Paris Cit\'{e}, F-91191 Gif-sur-Yvette, France\\
$^{13}$ Department of Physics, Federal University of Sergipe, S\~{a}o Cristov\~{a}o, SE, 49100-000, Brazil\\
$^{14}$ Institute of Theoretical Astrophysics, University of Oslo, P.O. Box 1029 Blindern, N-0315 Oslo, Norway\\
$^{15}$ Leiden Observatory, Leiden University, Niels Bohrweg 2, 2333 CA Leiden, The Netherlands\\
$^{16}$ INAF-IASF Milano, Via Alfonso Corti 12, I-20133 Milano, Italy\\
$^{17}$ Universit\'e PSL, Observatoire de Paris, Sorbonne Universit\'e, CNRS, LERMA, F-75014, Paris, France\\
$^{18}$ CEICO, Institute of Physics of the Czech Academy of Sciences, Na Slovance 2, Praha 8, Czech Republic\\
$^{19}$ Institut de Recherche en Astrophysique et Plan\'etologie (IRAP), Universit\'e de Toulouse, CNRS, UPS, CNES, 14 Av. Edouard Belin, F-31400 Toulouse, France\\
$^{20}$ Mullard Space Science Laboratory, University College London, Holmbury St Mary, Dorking, Surrey RH5 6NT, UK\\
$^{21}$ School of Physics and Astronomy, Queen Mary University of London, Mile End Road, London E1 4NS, UK\\
$^{22}$ Universit\'e St Joseph; UR EGFEM, Faculty of Sciences, Beirut, Lebanon\\
$^{23}$ Departamento de F\'isica, FCFM, Universidad de Chile, Blanco Encalada 2008, Santiago, Chile\\
$^{24}$ INAF-Osservatorio di Astrofisica e Scienza dello Spazio di Bologna, Via Piero Gobetti 93/3, I-40129 Bologna, Italy\\
$^{25}$ INAF-Osservatorio Astronomico di Padova, Via dell'Osservatorio 5, I-35122 Padova, Italy\\
$^{26}$ Universit\'e de Paris, CNRS, Astroparticule et Cosmologie, F-75006 Paris, France\\
$^{27}$ INFN-Sezione di Roma Tre, Via della Vasca Navale 84, I-00146, Roma, Italy\\
$^{28}$ Department of Mathematics and Physics, Roma Tre University, Via della Vasca Navale 84, I-00146 Rome, Italy\\
$^{29}$ INAF-Osservatorio Astronomico di Capodimonte, Via Moiariello 16, I-80131 Napoli, Italy\\
$^{30}$ Institut de F\'{i}sica d’Altes Energies (IFAE), The Barcelona Institute of Science and Technology, Campus UAB, 08193 Bellaterra (Barcelona), Spain\\
$^{31}$ Department of Physics "E. Pancini", University Federico II, Via Cinthia 6, I-80126, Napoli, Italy\\
$^{32}$ INFN section of Naples, Via Cinthia 6, I-80126, Napoli, Italy\\
$^{33}$ INAF-Osservatorio Astrofisico di Arcetri, Largo E. Fermi 5, I-50125, Firenze, Italy\\
$^{34}$ Dipartimento di Fisica e Astronomia, Universit\'a di Bologna, Via Gobetti 93/2, I-40129 Bologna, Italy\\
$^{35}$ Centre National d'Etudes Spatiales, Toulouse, France\\
$^{36}$ Institute for Astronomy, University of Edinburgh, Royal Observatory, Blackford Hill, Edinburgh EH9 3HJ, UK\\
$^{37}$ University of Nottingham, University Park, Nottingham NG7 2RD, UK\\
$^{38}$ European Space Agency/ESRIN, Largo Galileo Galilei 1, 00044 Frascati, Roma, Italy\\
$^{39}$ ESAC/ESA, Camino Bajo del Castillo, s/n., Urb. Villafranca del Castillo, 28692 Villanueva de la Ca\~nada, Madrid, Spain\\
$^{40}$ Aix-Marseille Univ, CNRS, CNES, LAM, Marseille, France\\
$^{41}$ Universit\'e Paris-Saclay, CNRS, Institut d'astrophysique spatiale, 91405, Orsay, France\\
$^{42}$ Department of Astronomy, University of Geneva, ch. d'\'Ecogia 16, CH-1290 Versoix, Switzerland\\
$^{43}$ INFN-Padova, Via Marzolo 8, I-35131 Padova, Italy\\
$^{44}$ Department of Physics \& Astronomy, University of Sussex, Brighton BN1 9QH, UK\\
$^{45}$ INAF-Osservatorio Astronomico di Trieste, Via G. B. Tiepolo 11, I-34131 Trieste, Italy\\
$^{46}$ Istituto Nazionale di Astrofisica (INAF) - Osservatorio di Astrofisica e Scienza dello Spazio (OAS), Via Gobetti 93/3, I-40127 Bologna, Italy\\
$^{47}$ Istituto Nazionale di Fisica Nucleare, Sezione di Bologna, Via Irnerio 46, I-40126 Bologna, Italy\\
$^{48}$ Universit\"ats-Sternwarte M\"unchen, Fakult\"at f\"ur Physik, Ludwig-Maximilians-Universit\"at M\"unchen, Scheinerstrasse 1, 81679 M\"unchen, Germany\\
$^{49}$ Max Planck Institute for Extraterrestrial Physics, Giessenbachstr. 1, D-85748 Garching, Germany\\
$^{50}$ INAF-Osservatorio Astronomico di Brera, Via Brera 28, I-20122 Milano, Italy\\
$^{51}$ Jet Propulsion Laboratory, California Institute of Technology, 4800 Oak Grove Drive, Pasadena, CA, 91109, USA\\
$^{52}$ von Hoerner \& Sulger GmbH, Schlo{\ss}Platz 8, D-68723 Schwetzingen, Germany\\
$^{53}$ Max-Planck-Institut f\"ur Astronomie, K\"onigstuhl 17, D-69117 Heidelberg, Germany\\
$^{54}$ Aix-Marseille Univ, CNRS/IN2P3, CPPM, Marseille, France\\
$^{55}$ Institut d'Astrophysique de Paris, 98bis Boulevard Arago, F-75014, Paris, France\\
$^{56}$ Universit\'e de Gen\`eve, D\'epartement de Physique Th\'eorique and Centre for Astroparticle Physics, 24 quai Ernest-Ansermet, CH-1211 Gen\`eve 4, Switzerland\\
$^{57}$ Department of Physics and Helsinki Institute of Physics, Gustaf H\"allstr\"omin katu 2, 00014 University of Helsinki, Finland\\
$^{58}$ NOVA optical infrared instrumentation group at ASTRON, Oude Hoogeveensedijk 4, 7991PD, Dwingeloo, The Netherlands\\
$^{59}$ Argelander-Institut f\"ur Astronomie, Universit\"at Bonn, Auf dem H\"ugel 71, 53121 Bonn, Germany\\
$^{60}$ Institute for Computational Cosmology, Department of Physics, Durham University, South Road, Durham, DH1 3LE, UK\\
$^{61}$ Observatoire de Sauverny, Ecole Polytechnique F\'ed\'erale de Lau- sanne, CH-1290 Versoix, Switzerland\\
$^{62}$ CEA Saclay, DFR/IRFU, Service d'Astrophysique, Bat. 709, 91191 Gif-sur-Yvette, France\\
$^{63}$ INFN-Bologna, Via Irnerio 46, I-40126 Bologna, Italy\\
$^{64}$ Department of Physics and Astronomy, University of Aarhus, Ny Munkegade 120, DK–8000 Aarhus C, Denmark\\
$^{65}$ Space Science Data Center, Italian Space Agency, via del Politecnico snc, 00133 Roma, Italy\\
$^{66}$ Institute of Space Science, Bucharest, Ro-077125, Romania\\
$^{67}$ Dipartimento di Fisica e Astronomia “G.Galilei", Universit\'a di Padova, Via Marzolo 8, I-35131 Padova, Italy\\
$^{68}$ INFN-Sezione di Bologna, Viale Berti Pichat 6/2, I-40127 Bologna, Italy\\
$^{69}$ Instituto de Astrof\'isica e Ci\^encias do Espa\c{c}o, Faculdade de Ci\^encias, Universidade de Lisboa, Tapada da Ajuda, PT-1349-018 Lisboa, Portugal\\
$^{70}$ Departamento de F\'isica, Faculdade de Ci\^encias, Universidade de Lisboa, Edif\'icio C8, Campo Grande, PT1749-016 Lisboa, Portugal\\
$^{71}$ Universidad Polit\'ecnica de Cartagena, Departamento de Electr\'onica y Tecnolog\'ia de Computadoras, 30202 Cartagena, Spain\\
$^{72}$ Kapteyn Astronomical Institute, University of Groningen, PO Box 800, 9700 AV Groningen, The Netherlands\\
$^{73}$ Infrared Processing and Analysis Center, California Institute of Technology, Pasadena, CA 91125, USA\\
}

\date{}

\authorrunning{M.\ Martinelli et al.}
\titlerunning{Euclid: impact of nonlinear prescriptions on cosmological parameter estimation}

 
  \abstract{
  {Upcoming surveys  will map the growth of large-scale structure with unprecented precision, improving our understanding of the dark sector of the Universe. Unfortunately, much of the 
   cosmological information is encoded by the small scales, where the clustering of dark matter and the effects of astrophysical feedback processes are not fully understood.}
  {This can bias the estimates of cosmological parameters, which we study here for a joint analysis of mock \Euclid cosmic shear and \Planck cosmic microwave background data.}
  {We use different implementations for the modelling of the signal on small scales and find that they result in significantly different predictions. Moreover, the different nonlinear corrections lead to biased parameter estimates, especially when the analysis is extended into the highly nonlinear regime, with both the Hubble constant, $H_0$, and the clustering amplitude, $\sigma_8$, affected the most. }
  {Improvements in the modelling of nonlinear scales will therefore be needed if we are to resolve the current tension with more and better data. For a given prescription for the nonlinear power spectrum, using different corrections for baryon physics does not significantly impact the precision of \Euclid, but neglecting these correction does lead to large biases in the cosmological parameters.}
  {In order to extract precise and unbiased constraints on  cosmological parameters from \Euclid cosmic shear data, it is therefore essential to improve the accuracy of the recipes that account for nonlinear structure formation, as well as the modelling of the impact of astrophysical processes that redistribute the baryons.}
  
   }


   \maketitle
%

\section{Introduction}\label{sec:intro}

The next generation surveys (stage IV) of the cosmic large-scale structure will greatly improve both the amount and quality of data for cosmological investigations. For instance, in the coming decade the surveys carried out by \Euclid\footnote{\url{https://www.euclid-ec.org}}, the Vera Rubin C. Observatory\footnote{\url{https://www.lsst.org}}, and the
{\it Nancy Grace Roman} Space Telescope\footnote{\url{https://roman.gsfc.nasa.gov/}} will probe scales and redshifts that were previously inaccessible. The correct interpretation of such a large amount of high-quality data, however, also poses a challenge for our theoretical modelling. 

In this paper we explore the current modelling limitations for one of the most promising probes: cosmic shear, the measurement of the apparent distortions of galaxy shapes caused by the weak lensing (WL) effect of intervening matter between us and the distant sources \citep[see][for a recent review]{Kilbinger15}. It provides a direct way to trace the distribution of matter, and as a result it can be used to infer the total matter power spectrum, $P_{\delta\delta}(k,z)$. In contrast, the galaxy power spectrum, $P_{\rm gg}(k,z)$, estimated from the clustering of galaxies, depends on the galaxy bias and how galaxies occupy nonlinear structures.

A complication is that much of the constraining power of the cosmic shear signal relies on being able to interpret scales far into the nonlinear regime, corresponding to wavenumbers $k\approx7\, h\,{\rm Mpc}^{-1}$ \citep[e.g.][]{Huterer05,Semboloni11,Taylor:2018nrc}. On those scales, perturbations of the matter density field are no longer small, and linear theory cannot be used to predict the evolution of large-scale structures. 

There are theoretical approaches to predict clustering beyond the limit of linear theory, such as: standard perturbation theory (\citealt{Blas_2014}; see also \citet{BERNARDEAU20021} for a detailed review) that includes higher-order terms; renormalised perturbation theory \citep{PhysRevD.73.063519,Crocce2012,Blas2016}; response functions \citep{NISHIMICHI2016247}; effective field theory \citep{Baumann2012}; or the reaction method \citep{Cataneo:2018cic}.
These methods are able to achieve accuracies on power spectrum predictions of 
$\approx1\%$ with respect to numerical simulations, up to scales $k\lesssim0.3\,h\,{\rm Mpc}^{-1}$ \citep[see e.g.][]{Foreman:2015uva,Beutler:2016ixs,DAmico:2019fhj}.  This is sufficient
for modelling the mildly nonlinear regime, where the baryonic acoustic oscillation peak is located, but these techniques cannot  be used to predict the signal in the highly non-linear regime that WL analyses will probe. 

The common approach to model the nonlinear part of the power spectrum instead relies on fitting formulae determined from comparisons to $N$-body simulations of cold dark matter particles \citep[e.g. \texttt{Halofit};][]{10.1046/j.1365-8711.2003.06503.x}. While very economical in terms of CPU time, these fitting formulae have a limited range of applicability, because the simulations they are based on assume a specific model -- usually $\Lambda$CDM or minimal extensions with a constant dark energy equation of state parameter allowed to be different from $-1$. Thus, applying these corrections to models outside the range constrained by the simulations, e.g.\ a more general dark energy fluid, may lead to biased results \citep[see e.g.][]{casariniFisher,Seo2011}.

Recently, so-called `emulators' -- based on a large suite of simulations, such as the \texttt{Coyote Universe} \citep{coyote1,coyote2,coyote3,coyote4}, the \texttt{Mira Universe} \citep{mira1, mira2}, and the \texttt{Euclid Emulator Project} \citep{EuclidEmulator} -- have been proposed as an alternative to fitting formulae. Emulators interpolate between high-resolution simulation runs at key points (nodes) in the cosmological parameter space. The main advantage of an emulator with respect to a fitting formula is that it does not degrade the accuracy of the corrections over the parameter space sampled by the simulations, such as the range in redshift and scales.

Another way to predict the matter power spectrum on small scales is provided 
by \texttt{HMCode}\footnote{\url{https://github.com/alexander-mead/HMcode}} \citep{Mead2015}, which is based on the analytical halo model \citep{Peacock:2000qk,Seljak:2000gq,Cooray2002}, and tuned to match the \texttt{Coyote Extended Emulator} \citep{coyote4} results. It has subsequently been improved to include effects of neutrinos, chameleon and Vainshtein screened models, and dynamical dark energy \citep{Mead2016}
\footnote{During the preparation of this work an update of \texttt{HMCode} was published, as described in \citet{Mead:2020vgs}.}.

All these methods, besides the declared quality of the method itself (for example the accuracy of the fitting formula, or the accuracy of the interpolation of the emulators), also depend on the quality of the simulations on which they are based. Restricting ourselves to just the nonlinear clustering, the agreement between purely dark matter simulations is limited by the size of the simulated volume, the number of particles employed in the simulation and the choice of initial conditions \citep[see e.g.][]{casariniSV,Schneider:2015wta}. Hence, part of the differences between the methods described above may be attributed to the simulation parameters on which they are based (e.g.\ when dimension and resolution are insufficient) rather than the methods themselves.

Moreover, our ability to extract cosmological information from WL measurements on small scales is limited further by baryonic feedback processes \citep{Semboloni11}, because gas cooling, star formation, galactic winds, supernova explosions, and feedback from active galactic nuclei modify the expected distribution of matter on small scales \citep{Jing2005,Rudd2007,casariniBar1,vanDaalen:2011xb,casariniBar2,2018MNRAS.478.1305C,2020MNRAS.492.2285D}. Accurate predictions of the matter power spectrum on those scales require hydrodynamical simulations that not only need to reproduce the nonlinear clustering of cold dark matter particles, but also should reliably describe the baryonic component. Such hydrodynamical simulations are much more demanding in terms of computational resources, and the impact of baryonic feedback extracted from hydrodynamical runs has to be modelled and incorporated in the reconstruction of the cosmic shear signal \citep[see e.g.][for a method to include the impact of feedback in the data analysis pipeline]{Schneider:2015wta,Schneider:2015yka}.

Given the cost of simulating large volumes with high-resolution for every necessary point in the parameter space for each specific cosmological model, we are particularly interested in techniques that can drastically reduce the number of simulations \citep[see][]{Linder2005, Francis2007, casariniPK1}. As an example, in this work we use the \texttt{PKequal}\footnote{\url{https://github.com/luciano-casarini/PKequal}} method that allows us to determine the nonlinear power spectrum of a dynamical dark energy model at a particular redshift with an ensemble of nonlinear spectra of simpler constant $w$ models \citep{casariniPK1,casariniPK2}.

In this paper, we investigate the impact of different implementations of the nonlinear corrections and baryonic feedback prescriptions on cosmological parameter estimation, adopting cosmic shear measurements from \Euclid \citep{Laureijs11} as our baseline. Notice that we do not investigate here other effects that might affect the parameter estimation pipeline, such as the common assumption of a Gaussian likelihood, which is only an approximation. The paper is organised as follows. After describing the various nonlinear recipes in \Cref{sec:NLrecipes}, in \Cref{sec:euspecs} we summarise the \Euclid specifications relevant for our analysis. As a first example of the impact of nonlinear prescriptions,  we assess in \Cref{sec:FoMeffect} the constraining power of the \Euclid survey on the dark energy parameters, here assumed to be described by the so-called CPL parameterisation \citep{Chevallier:2000qy,Linder:2002et}. In \Cref{sec:mcmcbias} we quantify the shifts in cosmological parameters when a wrong correction pipeline is used, focusing on the combination of {\it Planck} and mock \Euclid\ data. In \Cref{sec:bary} we examine the impact of baryonic feedback. 

\section{Available nonlinear prescriptions}\label{sec:NLrecipes}

In this section we describe the techniques that we use in this study to compute the matter power spectrum in the deeply nonlinear regime.

\subsection{\texttt{Halofit}}
One of the first widely used prescriptions to model the nonlinear part of the power spectrum, called \texttt{Halofit}, was developed by \citet{10.1046/j.1365-8711.2003.06503.x}. The authors measured the nonlinear evolution of the matter power spectra using a large library of cosmological $N$-body simulations with power-law initial spectra \citep{Jenkins1998}. 

The \texttt{Halofit} approach is based on the halo model \citep{Peacock:2000qk,Seljak:2000gq, Ma2000}, in which the density field is described in terms of the distribution of isolated dark matter haloes. The correlations in the field are assumed to arise from the clustering of haloes with respect to each other on large scales, and through the clustering of dark matter particles within the same halo on small scales. The total nonlinear power spectrum, $P_{\rm NL}(k)$, can then be decomposed into
\begin{equation}
P_{\rm NL}(k)=P_{\rm Q}(k)+P_{\rm H}(k)\,,
\end{equation}
where $P_{\rm Q}(k)$ is the quasi-linear term related to the large-scale contribution to the power spectrum, and $P_{\rm H}(k)$ describes the contribution from the self-correlation of haloes. These terms are also known as the 2-halo and the 1-halo term, respectively, and we discuss them in this order below.

\citet{Seljak:2000gq,Ma2000,Scoccimarro2001} proposed to use linear theory filtered by the effective window that corresponds to the distribution of haloes as a function of mass, $n(M)$, convolved with their density profiles, $\tilde{\rho}(k,M)$, and a prescription for their bias with respect to the underlying mass field, $b_{\rm H}(M)$. The quasi-linear term can then be expressed as
\begin{equation}
P_{\rm Q}(k)=P_{\rm L}(k)\left[\frac{1}{\bar{\rho}}\int {\rm d}M\,b_{\rm H}(M)n(M)\tilde{\rho}(k,M)\right]^2\, ,
\end{equation}
with $\bar{\rho}$ the homogeneous background matter density,
and $P_{\rm L}(k)$ the linear power spectrum.

A simpler approach was proposed by \citet{Peacock:2000qk}, who assumed that the quasi-linear term corresponds to pure linear theory, $P_{\rm Q}(k)=P_{\rm L}(k)$. However, quasi-linear effects must modify the relative correlations of haloes away from linear theory, irrespective of the allowance made for the finite size of the haloes (see \citealt{10.1046/j.1365-8711.2003.06503.x}, and references therein). \texttt{Halofit} takes then an empirical approach, allowing the quasi-linear term to depend on $n(M)$, and truncating its effects at small scales. If we define the dimensionless power spectrum as
\begin{equation}
\Delta^2(k)\equiv \frac{k^3}{2\pi^2}P(k)\,,
\end{equation}
the quasi-linear term in \texttt{Halofit} is given by
\begin{equation}
\Delta_{\rm Q}^2(k)=\Delta_{\rm L}^2(k)\frac{[1+\Delta_{\rm
    L}^2(k)]^{\beta_n}}{1+\alpha_n\Delta_{\rm L}^2(k)}\,{\rm e}^{-f(y)}\,,
    \label{eq.P_QHalofit}
\end{equation}
where $y\equiv k/k_{\sigma}$, $k_{\sigma}$ is a nonlinear wavenumber related to the spherical collapse model \citep{Press1974,Sheth1999,Jenkins2001}, $\alpha_n$ and $\beta_n$ are coefficients sensitive to the input linear spectrum, and $f(y)=y/4+y^2/8$ governs the decay rate at small scales.

To describe the clustering of matter on small scales, we need a description for $P_{\rm H}$, which is given by
\begin{equation}
P_{\rm H}(k)=\frac{1}{\bar{\rho}^2(2\pi)^3}\int {\rm d}M\,n(M)|\tilde{\rho}(k,M)|^2\,.
\end{equation}
In order to model this term, we can use an expression that looks like a shot-noise spectrum on large scales, but progressively vanishes on small scales by the filtering effects of halo profiles and the mass function. A good candidate is
\begin{equation}\label{eq.P_H}
{\Delta_{\rm H}^2}'(k)=\frac{a_n y^3}{1+b_ny+c_ny^{3-\gamma_n}}\,,
\end{equation}
where $a_n,\,b_n,\,c_n$, and $\gamma_n$ are dimensionless numbers that depend on the input spectrum. 

\citet{Cooray2002}, however, showed that with this expression the halo model 
disagrees with low-order perturbation theory in some cases. To solve this, \texttt{Halofit} modifies \Cref{eq.P_H} to obtain a spectrum steeper than Poisson on the largest scales
\begin{equation}
\Delta_{\rm H}^2(k)=\frac{{\Delta_{\rm H}^{2}}'(k)}{1+\mu_n y^{-1}+\nu_n y^{-2}}\,,
\end{equation}
where $\mu_n$ and $\nu_n$ are, again, coefficients that depend on the input spectrum. \citet{10.1046/j.1365-8711.2003.06503.x} showed that \texttt{Halofit} is able to reproduce the measurements from simulations more accurately, and down to smaller scales, than the
halo model.

\subsection{\texttt{Halofit} with Bird and Takahashi corrections}
All the coefficients used in the \texttt{Halofit} recipe were determined by \citet{10.1046/j.1365-8711.2003.06503.x} from a fit to cold dark matter simulations in boxes of lengths of  256$\,h^{-1}\,\mathrm{Mpc}$ containing $256^3$ particles \citep{Jenkins1998}. As a consequence of the relatively large particle mass, \texttt{Halofit} may not be suitable if  we want to test cosmologies with massive neutrinos, or go down to very small  scales, where the impact of baryonic interactions is non-negligible. Moreover, the limited simulation volume results in  large sample variance \citep[see][]{casariniSV,Schneider:2015wta}, and thus may lead to inaccurate results even for a $\Lambda$CDM cosmology \citep{WhiteVale2004,casariniPK1,Hilbert2009,coyote1,casariniBar2}. Finally, as the simulations were performed for the standard cosmological model, using \texttt{Halofit} with a dark energy equation of state $w\neq-1$ may yield an incorrect estimate of the power spectrum \citep{casariniFisher,Seo2011}.

To address these limitations, \citet{Bird2012} investigated the impact of massive neutrinos, and performed several $N$-body simulations of the matter power spectrum incorporating massive neutrinos with masses between $0.15$ and $0.6\,\mathrm{eV}$. They focussed on nonlinear scales below $10\,h\,\mathrm{Mpc}^{-1}$ at $z<3$,
and extended the \texttt{Halofit} approximation to account for massive neutrinos.
They found that in the strongly nonlinear regime \texttt{Halofit} over-predicts the suppression due to the free-streaming of the neutrinos. In particular, the asymptotic behaviour of the nonlinear term in \texttt{Halofit} is given by $\Delta_{\rm H}^2 \sim y^{\gamma_n}$, and therefore \citet{Bird2012} adjusted $\gamma_n$ to their $\Lambda$CDM simulations with massive neutrinos. Moreover, they modified the nonlinear power spectrum with the ansatz
\begin{equation}
(\Delta_{\rm NL}^{\nu})^2 = \Delta_{\rm NL}^2(1+Q_\nu),
\end{equation}
with
\begin{equation}
Q_\nu = \frac{l f_\nu}{1+k^3m},
\end{equation}
where $f_\nu=\Omega_\nu/\Omega_{\rm m}$ is the ratio between the neutrino and total matter energy densities, and $l$ and $m$ are fitted to the simulations. They also modified  \Cref{eq.P_QHalofit} to
\begin{equation}
\Delta_{\rm Q}^2(k)=\Delta_{\rm L}^2(k)\frac{[1+\tilde{\Delta}_{\rm
    L}^2(k)]^{\tilde{\beta}_n}}{1+\alpha_n\tilde{\Delta}_{\rm L}^2(k)}\,{\rm e}^{-f(y)}\,,
\end{equation}
with
\begin{align}
\tilde{\Delta}^2_{\rm L} &= \Delta_{\rm L}^2\left(1+\frac{pf_\nu k^2}{1+1.5k^2}\right)\,,\\
\tilde{\beta}_n &= \beta_n +f_\nu(r+n^2s)\,,
\end{align}
where $p,r,$ and $s$ are fit to the simulations. 

Another important improvement to the original \texttt{Halofit} was introduced in \citet{Takahashi2012}, who updated the fitting parameters using high-resolution $N$-body simulations with box sizes of $L=300-2000\,h^{-1}\,\mathrm{Mpc}$ with $n_{\rm p}=1024^3$ particles each, for 16 cosmological models around the best fitted  cosmological parameters from WMAP data \citep{2013ApJS..208...19H}, including dark energy models with a constant equation of state. This revised version of \texttt{Halofit} provides an accurate prediction of the nonlinear matter power spectrum down to $k\sim 30\,h\,\mathrm{Mpc}^{-1}$ and up to $z\geq 3$ with an accuracy $\sim$ 5 -- 10 \%. In the remainder of this paper we refer to the nonlinear prescription that includes the improvements 
from \citet{Takahashi2012} and \citet{Bird2012} as \texttt{Halofit}, for simplicity. 

\subsection{\texttt{Halofit} with \texttt{PKequal}}
One of the limitations of the standard \texttt{Halofit} approach, even after the corrections by \citet{Bird2012} and \citet{Takahashi2012} are considered, is that it is based on a fit to $N$-body simulations with a constant value for the dark energy equation of state. However, given the precision of stage IV surveys, we are particularly interested in determining whether the data prefer an evolving dark energy equation of state. To avoid biases in our nonlinear predictions one could run $N$-body simulations that include a time dependence for $w$, such as the CPL parameterisation \citep{Chevallier:2000qy,Linder:2002et} given by \begin{equation}\label{eq:cpl}
    w(a)=w_0+w_a(1-a)\,.
\end{equation}

This approach implies the need for significant computational resources. Another option, however, is to map the nonlinear power spectra of dark energy models with a constant equation of state to those with a time varying one. In this work we consider the \texttt{PKequal} code \citep{casariniPK2}, that implements the spectral equivalence from \citet{casariniPK1}. \cite{Francis2007} showed how predictions for constant $w$ models at $z=0$ can be related to the power spectra of cosmologies with an evolving equation of state $w(a)$ given by \Cref{eq:cpl} with an accuracy $\sim 0.5$\% up to $k\simeq 1\, h\, \mathrm{Mpc}^{-1}$. The \texttt{PKequal} technique achieves this precision also at $z>0$ for a general equation of state $w=w(a)$ by imposing the equivalence of the distance to the last scattering surface and requiring that the amplitudes of the density fluctuations at the redshift of interest are the same. For a given set of values of $w_0$ and $w_a$ these two conditions yield at each $z$ a unique value of $w_{\rm eq}$ and $\sigma_{8,{\rm eq}}$ for the constant $w$ model.

The performance of this method has been tested for several dark energy models \citep{casariniPK1,casariniPKtest}, and also in the presence of gas cooling, star formation and SN feedback \citep{casariniBar1}. These studies find differences in power spectra between the mapped dynamical dark energy models and the ensemble of equivalent constant $w$  models that are within $1\, \%$ up to $k\simeq 2$ -- $3\, h\, \mathrm{Mpc}^{-1}$. With this method it is possible to extend both emulators \citep[see][]{casariniPK2} and fitting formulae (as in this work) to dynamical dark energy models if they are valid for constant $w$ models.

\subsection{\texttt{HMCode}}
An alternative approach to predict the nonlinear matter power spectrum, called \texttt{HMCode}, was proposed by \citet{Mead2015}.
It introduces physically motivated free parameters into the halo model formalism, instead of using empirical fitting functions.
\citet{Mead2015} fit these to $N$-body simulations with box sizes  $L=90$ -- $1300\,h^{-1}\,\mathrm{Mpc}$ and $n_{\rm p}=512^3$ -- $1024^3$ particles for a variety of $\Lambda$CDM and $w$CDM models \citep{coyote1}. \texttt{HMCode} also accounts for the effects of baryonic feedback on the power spectrum by fitting the halo model to hydrodynamical simulations that include gas cooling, star formation, as well as supernova and AGN feedback \citep{Schaye2010,vanDaalen:2011xb}.

In \citet{Mead2016} \texttt{HMCode} was updated to account for massive neutrinos, chameleon and Vainshtein screening mechanisms, as well as  evolving dark energy equations of state described by the CPL parameterisation. Throughout the rest of the paper we will consider this latest version when referring to \texttt{HMCode}. We note, however, that the \texttt{PKequal} approach can also be used with the original \texttt{HMCode} by applying it to the equivalent $\{w_{\rm eq},0\}$ constant $w$ model at any redshift $z$, and imposing the same $\sigma_8(z)$. This yields the prediction for a given $\{w_0,w_a\}$ CPL model, with an accuracy similar to that of the fit and the simulations on which \texttt{HMCode} itself is based.

\subsection{Comparison}
Before we study how the different methods affect the inference of cosmological parameters and related confidence regions, we show
by how much the power spectra differ as a function of the wavenumber $k$ for the prescriptions described above. We note that theoretical approaches based on perturbation theory or the effective field theory of large-scale structure do not yet provide accurate power spectra for the small scales we consider here. 

\begin{figure*}
\centering
\includegraphics[scale=0.50]{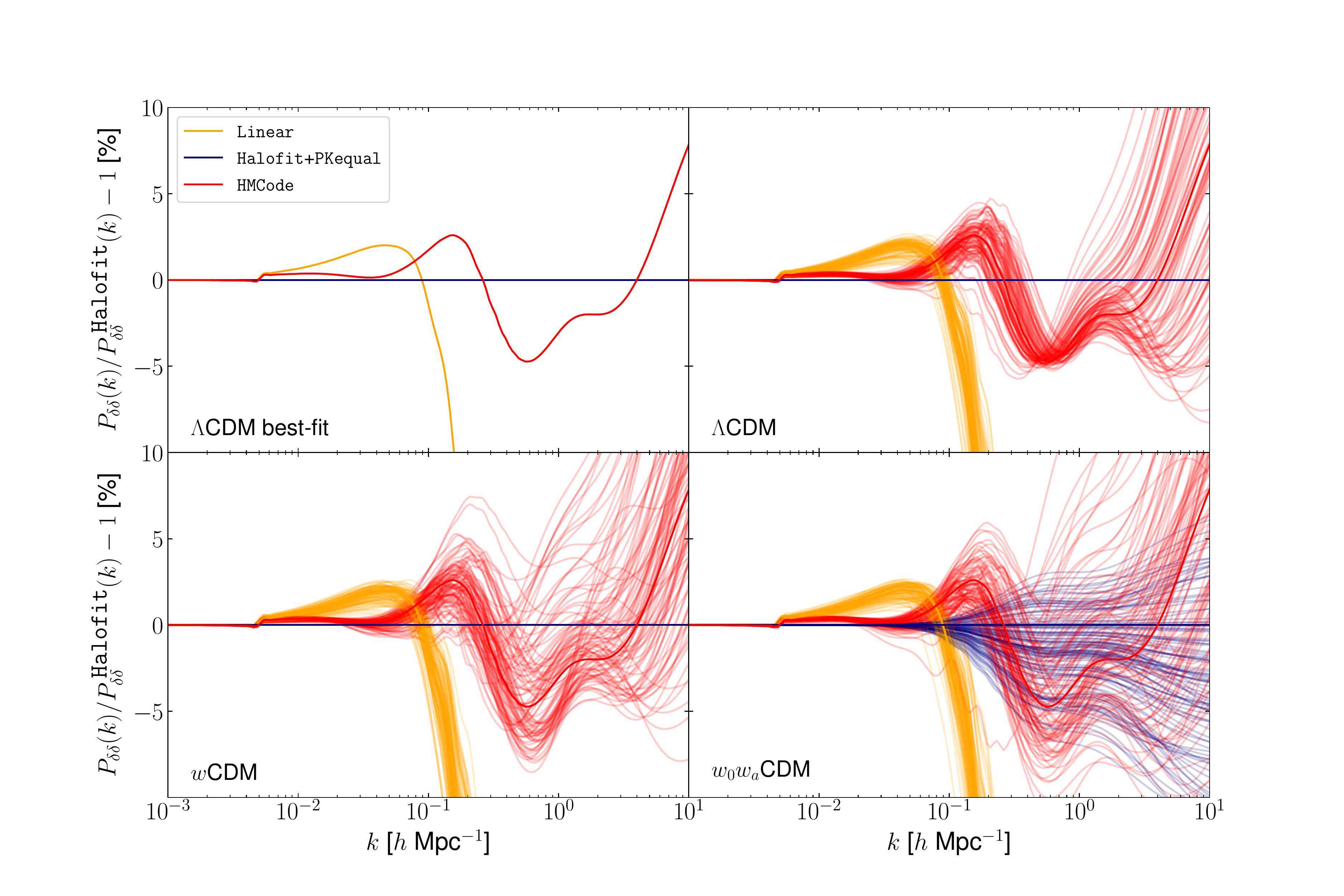}
\caption{Relative difference between the various predicted matter power spectra and the \texttt{Halofit} prediction at $z=0$ for
different cosmologies. {\it Top-left panel:} Using the mean  values of the $\Lambda$CDM parameters given by \textit{Planck} 2018 \citep{Aghanim:2018eyx}. {\it Top-right panel:} Differences when the $\Lambda$CDM parameters are  drawn from a 5-dimensional Gaussian distribution with the diagonal given by 5 times the $1\sigma$ uncertainty quoted by \textit{Planck} TTTEEE+lowE+lensing+BAO 2018. {\it Bottom-left panel:} Differences when also the constant dark energy equation of state is drawn from a Gaussian distribution with a dispersion of $0.3$. {\it Bottom-right panel:} Difference for evolving dark energy models where, alongside $w_0$, also $w_a$ is Gaussian distributed with a dispersion of $1.0$. In the bottom panels $w_0$ and $w_a$ (if present) are always chosen in such a way that $w(z)<-1/3.$}
\label{fig:pk_ratio}
\end{figure*}

In \Cref{fig:pk_ratio} we show the relative difference between the 
various predictions and the one from \texttt{Halofit}. The power spectra were computed using \texttt{CAMB} \citep{Lewis:1999bs} at redshift 0, where discrepancies are most pronounced. Hence, these represent a worst case scenario, as WL probes mostly intermediate redshifts. The upper-left panel shows the relative differences between the linear power spectrum (orange), \texttt{Halofit} with \texttt{PKequal} (blue), and \texttt{HMCode} (red) for the $\Lambda$CDM model using the \textit{Planck} TTTEEE+lowE+lensing+BAO 2018 mean values for the various cosmological parameters \citep{Aghanim:2018eyx}. The linear power spectrum diverges from \texttt{Halofit} for $k>0.1\,h\, \mathrm{Mpc}^{-1}$, while the disagreement between \texttt{Halofit} and \texttt{HMCode} is below 7.5\% for $k<10\,h\,\mathrm{Mpc}^{-1}$.

The upper-right panel shows the variation on small scales when 
we consider parameters around the mean values of $\Lambda$CDM. For this comparison we generated 100 power spectra with cosmological parameters drawn from a 5-dimensional Gaussian distribution with the diagonal given by 5 times the $1\sigma$ uncertainty quoted by \textit{Planck} TTTEEE+lowE+lensing+BAO 2018\footnote{Note that we fix the value of $\tau$ to its mean value; therefore we are left with a 5-dimensional Gaussian instead of the 6-dimensional one that would correspond to $\Lambda$CDM.}.
These results highlight  that the discrepancy between \texttt{Halofit} and \texttt{HMCode} can be larger than 10\%  for $k>3\,h\,\mathrm{Mpc}^{-1}$.

The bottom-left panel considers models beyond $\Lambda$CDM, allowing for a constant equation of state parameter, $w$, that can differ from $-1$. The different lines 
correspond to power spectra when we draw parameters from a 6-dimensional Gaussian distribution where we adopt a dispersion of $0.3$ for $w$ (but we do require $w<-1/3$). The bands showing the discrepancy between the different nonlinear corrections increase slightly (a $10\%$ discrepancy is reached at scales of $k=1$ -- $2\,h\,\mathrm{Mpc}^{-1}$), but the overall shape remains the same.

Finally, in the bottom-right panel we consider dynamical dark energy models with a dark energy equation of state given by \Cref{eq:cpl}. In this case we add $w_a$ (so that we draw parameters from a 7-dimensional Gaussian) with a dispersion of $1.0$ and we require the dark energy equation of state to be always smaller than $-1/3$, i.e.\ $w_0+w_a < -1/3$. The overall shape for the bands is the same for \texttt{HMCode} and the linear spectra, although the discrepancies are significantly larger (already 10\% at scales of $k=0.6\,h\,\mathrm{Mpc}^{-1}$). We also note that, since we allow $w_a$ to vary, there is a difference between \texttt{Halofit} and \texttt{Halofit}+\texttt{PKequal}.

\cite{mira2} presented an updated version of \texttt{COSMIC EMU} that includes massive neutrinos. Their high-resolution simulations are interpolated with an accuracy of $\sim 4 $\%.  \citet{mira2} compared their \texttt{COSMIC EMU} predictions with \texttt{Halofit}, and \texttt{HMCode}. When massive neutrinos are considered, the different approaches show differences of $ \sim 20$\% and $\sim 15$\% respectively in the power spectra for scales above $k=0.1\,{\rm Mpc}^{-1}$, indicating the need for further improvements.

\section{The Euclid Cosmic Shear survey}
\label{sec:euspecs}

Our aim is to investigate how the expected constraints on cosmological parameters from \Euclid data depend on the recipe that is used to predict the matter power spectrum on nonlinear scales, although we note that our finding are also relevant for other stage IV experiments. \Euclid is an M-class mission of the European Space Agency \citep{Laureijs11} that will carry out a spectroscopic and a photometric survey of galaxies over an area of 15\,000~deg$^2$.
The cosmic shear measurements use high-quality imaging at optical wavelengths, supported by multi-band optical ground-based photometry 
and near-infrared observations by \Euclid. The telescope is designed so that (residual) instrumental sources of bias in the observed cosmic shear signal are subdominant compared to the statistical uncertainties \citep[e.g.][]{Cropper13,Paykari20}. However, to achieve its objectives, it is essential that the signal can be accurately predicted in the nonlinear regime. Although this is also relevant to fully exploit the data from the clustering of galaxies and the cross-correlations with the lensing signal, we focus on the cosmic shear case in this paper and defer a more comprehensive study to future work.

We adopt the baseline specifications for the \Euclid data, which are 
described in \citet[][hereafter EC19]{IST:paper1}. The redshift distribution of the sources is given by
\begin{equation}
    n(z)\propto\left(\frac{z}{z_0}\right)^2
    \exp\left[-\left(\frac{z}{z_0}\right)^{3/2}\right],
\end{equation}
with $z_0=0.9/\sqrt{2}$, resulting in a mean redshift of $\langle z\rangle=0.96$. The sample is divided into 10 equi-populated redshift bins $n_i(z)$ (with the index $i$ indicating the tomographic bin). We assume an average number density of galaxies with precise shape measurements of $\bar{n}_g=30\ {\rm arcmin}^{-2}$. To capture the noise arising from the intrinsic galaxy ellipticities we adopt a dispersion of $\sigma_\epsilon=0.21$ for each of the two ellipticity components. We also use the same approach as \citetalias{IST:paper1} to model the data covariances and to compute the WL power spectrum, defined as 
\begin{equation}\label{eq:cells}
 C^{\epsilon\epsilon}_{ij}(\ell) = c\int{\rm d}z\,\frac{W^\epsilon_i(z)W^\epsilon_j(z)}{H(z)r^2(z)}P_{\delta\delta}\left[\frac{\ell+1/2}{r(z)},z\right],
\end{equation}
where $P_{\delta\delta}$ is the nonlinear matter power spectrum, $r(z)$ is the comoving distance to redshift $z$, and the window function $W^\epsilon_i$ is defined as
\begin{equation}
    W_i^\epsilon(z) = W_i^\gamma(z)-\frac{\mathcal{A}_{\rm IA}\mathcal{C}_{\rm IA}\Omega_{{\rm m},0}\mathcal{F}_{\rm IA}(z)H(z)n_i(z)}{D(z)c},\label{eq:W_epsilon}
\end{equation}
with $n_i(z)$ normalised such that $\int{\rm d}z\,n_i(z)=1$. In \Cref{eq:W_epsilon}, the first term corresponds to the usual lensing kernel,
\begin{align}
    W^{\gamma}_{i}(z) = &
\frac{3}{2}\Omega_{{\rm m},0}\left(\frac{H_0}{c}\right)^2(1 + z) r(z) \nonumber \\
&\times \int_{z}^{z_{\rm max}}\mathrm d z^{\prime}\,n_i(z^{\prime})\left[ 1 - \frac{r(z)}{r(z^{\prime})} \right],
\end{align}
whilst the second term models the effect of intrinsic alignments (IA). Here $D(z)$ is the growth factor, ${\cal{C}}_{\rm IA}=0.0134$ is a constant so that the normalisation of the model $\mathcal A_{\rm IA}$ can be compared to the literature. To describe the dependence of the IA signal as a function of scale, redshift and galaxy luminosity, we adopt the extended nonlinear alignment model  \citep{iareview1}, and the function ${\cal{F}}_{\rm IA}(z)$ is given by
\begin{equation}
{\cal{F}}_{\rm IA}(z) = (1 + z)^{\eta_{\rm IA}}[\langle L \rangle(z)/L_{\star} (z)]^{\beta_{\rm IA}}\,,
\label{eq: effeia}
\end{equation}
where $\langle L \rangle(z)$ and $L_{\star}(z)$ are the redshift-dependent mean and a characteristic luminosity of source galaxies as computed from the luminosity function, respectively. 
The parameters $\eta_{\rm IA}$, $\beta_{\rm IA}$, and ${\cal{A}}_{\rm IA}$ are free parameters that can be determined observationally. We use $\{\mathcal A_{\rm IA},\eta_{\rm IA},\beta_{\rm IA}\}=\{1.72,-0.41,2.17\}$ as fiducial values, as was done in \citetalias{IST:paper1}, while they are allowed to vary in the parameter estimation.

Our approach for the modelling of IA is phenomenological, but we note that more physically motivated models have been proposed. For instance, \cite{Fortuna:2020vsz} used a halo model approach to link direct observations of IA to implications for cosmic shear. Finally, we note that the modelling of the IA signal is linked to nonlinear structure formation, but exploring this further is beyond the scope of this paper. 

\section{Impact of nonlinear corrections on forecasted constraints}
\label{sec:FoMeffect}

Given our desire to use measurements on small scales to estimate cosmological parameters, it is essential to assess how the different methods to model the power spectrum on highly nonlinear scales affect the results. In the following, we limit ourselves to a simple extension of the $\Lambda$CDM model: we assume that dark energy is dynamical with its equation of state parameterised according to \Cref{eq:cpl}, and that density perturbations for this component are well described by the parameterised post-Friedmann approach, which assumes that the dark energy field remains smooth with respect to matter at the scales of interest \citep{Hu:2007pj,Hu:2008zd,Fang:2008sn}.

To evaluate the impact of the different recipes for the nonlinear power spectrum on the final parameter estimation from \Euclid, we use the Fisher matrix approach \citepalias[see][for an extensive review of the methodology]{IST:paper1}. For the different methods we compute 
the `figure of merit' (FoM) for the parameters $w_0$ and $w_a$, where the FoM is defined as 
\begin{equation}
{\rm FoM} = \sqrt{{\rm det}\tilde{\tens F}_{w_0w_a}},
\end{equation}
and $\tilde{\tens F}_{w_0w_a}$ is the Fisher matrix marginalised over all the cosmological parameters except for $w_0$ and $w_a$. This allows us to examine whether degeneracies between cosmological parameters differ when switching from one method to the other. We can also quantify the biases in dark energy parameters and changes in the 
FoM, which captures the performance of \Euclid. 

The free parameters for this analysis are: the matter and baryon density parameters $\Omega_{\rm m,0}$ and $\Omega_{\rm b,0}$; the dark energy parameters, $w_0$ and $w_a$; the  spectral index of primordial perturbations, $n_{\rm s}$; the dimensionless Hubble parameter, $h$; and $\sigma_8$, i.e.\ the root mean square of present-day linearly evolved density fluctuations in spheres of $8\,h^{-1}\,\mathrm{Mpc}$ radius. The fiducial values of these parameters are listed in \Cref{tab:fiducialfisher}.

\begin{table}[!h]
\centering
\caption{Fiducial values for the cosmological parameters considered in the Fisher matrix analysis.}
\begin{tabular}{ccccccc} 
 \hline 
 $\Omega_{\rm m,0}$    &  $  \Omega_{\rm b,0}  $  & $ w_0  $ & $w_a$ & $ n_{\rm s}$ &$h$ &$\sigma_8$\\
\hline 
 $0.32$ & $ 0.05$ & $-1$ & $0$ & $0.960$ & $0.67$ & $0.816$ \\
\hline
\end{tabular} \\
\label{tab:fiducialfisher}
\end{table}

We use the specification of \Euclid  for WL observations that were detailed in \Cref{sec:euspecs} and compute the FoM from the WL power spectra $C_\ell^{\epsilon\epsilon}$ for different values of the maximum multipole, $\ell_{\rm max}$. We consider two different values for $\ell_{\rm max}$, namely $1500$ and $5000$, with the latter probing deeper into the nonlinear regime. Notice that here the same $\ell_{\max}$ applies to all redshift bins, thus leading to a different cut in scales ($k_{\rm max}$) for each bin. We investigate the difference between this analysis and one considering a $k_{\rm max}$ rather than a multipole cut in \Cref{sec:nlcut}. In both cases the minimum multipole is fixed to $\ell_{\rm min}=10$ following \citetalias{IST:paper1}. We note that the cuts we use here should be considered an approximation; in general there is no direct mapping between $\ell_{\rm max}$ and $k_{\rm max}$, and a more refined approach would be needed to convert a cut multipoles into a cut in wavenumber \citep{Taylor:2018nrc}. 

\begin{table}[!h]
\centering
\caption{FoM estimated using \Euclid specifications for WL with \texttt{Halofit}, \texttt{HMCode}, and \texttt{Halofit}+\texttt{PKequal}.}
\begin{tabular}{lccc} 
 \hline 
$\ell_{\rm max}$  & \texttt{Halofit} & \texttt{HMCode} & \texttt{Halofit}+\texttt{PKequal}  \\
\hline 
1500 & $23$  & $14$ & $16$ \\
5000 & $44$ & $34$ & $36$ \\
\hline 
\end{tabular}
\label{tab:FoMchange}

\end{table}

\Cref{tab:FoMchange} lists the resulting FoM values for our baseline $w_0w_a$CDM cosmology for the three different nonlinear recipes. 
The variation in the predicted FoMs is substantial, which is not that surprising given the differences we see in \Cref{fig:pk_ratio}; at large $k$ the differences exceed 10\% when the parameters are allowed to vary with respect to the fiducial model. 

The large variation is caused by two separate effects. Firstly, the fiducial $C_{\ell}^{\epsilon\epsilon}$ is obtained by integrating up to $k \simeq 30\, h\,{\rm Mpc}^{-1}$ and then inverted when computing the covariance matrix that enters the Fisher matrix forecast. Small differences can become large in the inversion process. Secondly, what is important is not so much the fiducial model itself, but rather the derivatives of the power spectrum with respect to the model parameters. Different nonlinear corrections predict different derivatives, thus leading to different Fisher matrix elements. 
This is supported by the fact that the FoM for \texttt{HMCode} and \texttt{Halofit}+\texttt{PKequal} are very similar. These two prescriptions explicitly take  deviations of $w_0$ and $w_a$ from the fiducial $\Lambda$CDM values into account, while this is not the case for the standard \texttt{Halofit}, which is designed to describe $\Lambda$CDM cosmologies and extended to cases where dark energy is described by a constant equation of state. As a consequence, the amplitude of the derivatives of the matter power spectrum with respect to $\{w_0,\,w_a\}$ are similar between \texttt{HMCode} and \texttt{Halofit}+\texttt{PKequal}, but different for \texttt{Halofit}. This explains why the FoM values in the third and fourth columns are so similar. 

It is also worth noting that the FoM is highest for \texttt{Halofit}, because in that case  $w_a$ impacts the power spectra at nonlinear scales through its effect on the linear matter power spectrum, while in the other two methods the nonlinear corrections are affected by this parameter as well. Overall, the nonlinear corrections dampen the derivatives with respect to $w_a$, leading to weaker constraints on this parameter and to a lower FoM. These results demonstrate that the FoM depends critically on the nonlinear model that is used, highlighting the need for (more) accurate prescriptions.

\section{Bias on cosmological parameter estimates}\label{sec:mcmcbias}

A major concern is that inaccuracies in the theoretical predictions on nonlinear scales translate into shifts in the inferred cosmological parameters. To quantify the impact of such biases, we create a mock data set of \Euclid WL observations, using the specifications listed in \Cref{sec:euspecs}. The fiducial cosmology that we use to generate these mock data is summarised in \Cref{tab:fiducialmcmc}, where 
 $A_s$ is the amplitude of the primordial power spectrum, $\omega_{{\rm b},0}=\Omega_{{\rm b},0}h^2$, $\omega_{{\rm c},0}=\Omega_{{\rm c},0}h^2$ where $\Omega_{{\rm c},0}$ is the present time density parameter for cold dark matter.
In contrast to what was done in \Cref{sec:FoMeffect}, we assume here that the expansion history is provided by a dynamical dark energy, assuming $w_0=-0.9$ and $w_a=0.1$. This allows us to assess the impact of the different nonlinear recipes when a non-standard model is the true underlying cosmology. This is motivated by the fact that the recipes in \Cref{sec:NLrecipes} differ in how they account for such extensions of the standard cosmological model.

We adopt the nonlinear correction provided by \texttt{\texttt{Halofit}+PKequal} as the reference to which we compare the parameter estimates for the other recipes. We stress that we do not advocate the use of a particular prescription, but rather wish to quantify the shifts in the estimated cosmological parameters that may arise from using a different prediction.

\begin{table}[!htbp]
\centering
\caption{Fiducial values for the free cosmological parameters in the MCMC analysis. We model neutrinos assuming $2.0328$ ultra-relativistic species, and $1$ mass eigenstate with $m_\nu=0.06$ eV. The MCMC fiducial model implies $\sigma_8=0.786$, which corresponds to the value of the Fisher matrix analysis rescaled according to the new cosmology, i.e.\ replacing the cosmological constant with a dynamical dark energy with $w_0=-0.9$ and $w_a=0.1$. }
\begin{tabular}{ll} 
\hline 
Parameter symbol & Parameter value \\
\hline 
 $\omega_{{\rm b},0}$ & $0.02245$ \\
 $\omega_{{\rm c},0}$ & $ 0.12056$ \\
 $h$ & $0.67$ \\
 $\ln (10^{10}A_{\rm s})$ & $3.05836$ \\
 $ n_{\rm s}$ & $0.96$ \\
 $\tau$ & $0.06$ \\
 $w_0$ & $-0.9$ \\
 $w_a$ & $0.1$ \\
\hline
\end{tabular}
\label{tab:fiducialmcmc}
\end{table}

\begin{figure*}[!htbp]
\centering
\begin{tabular}{cc}
\includegraphics[width=0.45\textwidth]{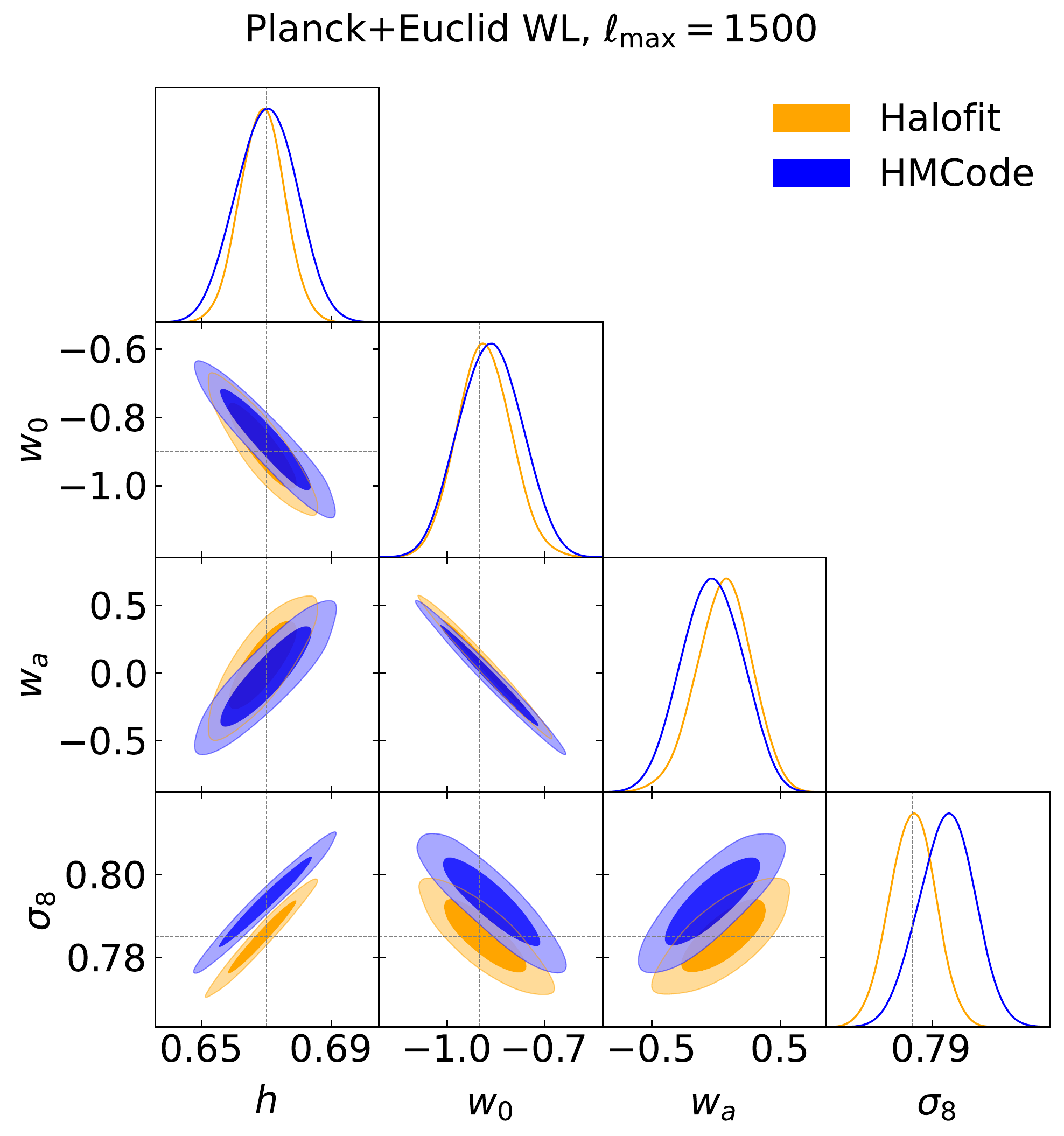}&
\includegraphics[width=0.45\textwidth]{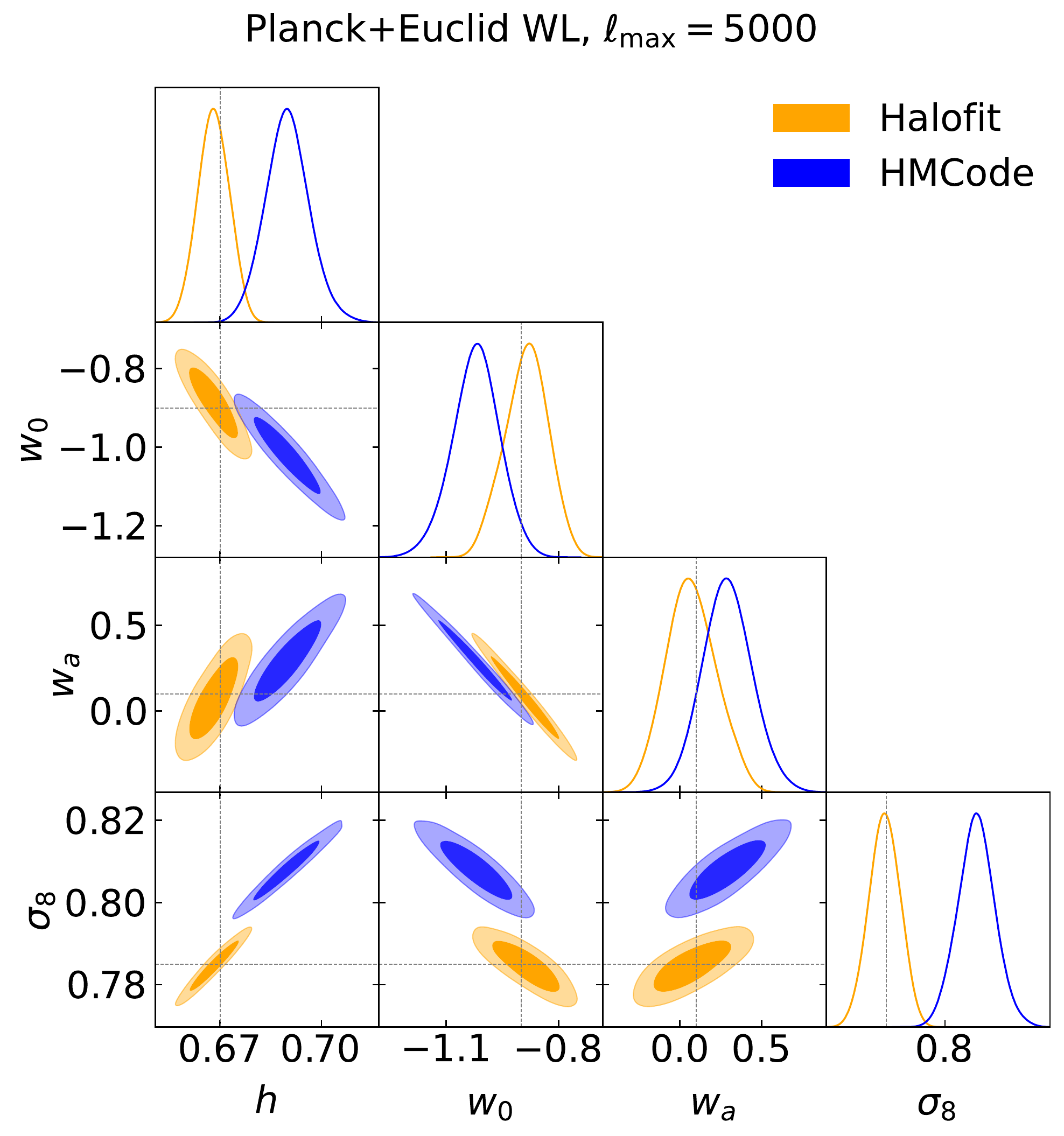}
\end{tabular}

\caption{One dimensional posterior distributions, and $68\%$ and $95\%$ confidence level marginalised contours for the dark energy parameters ($w_0$ and $w_a$) and the parameters $h$
and $\sigma_8$. The left panel refers to the $\ell_{\rm max}=1500$ case, while the right panel uses measurements deeper into the nonlinear regime, with $\ell_{\rm max}=5000$. The mock data of \Euclid cosmic shear assume \texttt{\texttt{Halofit}+PKequal} nonlinear corrections as reference, while the parameter estimation is performed with either \texttt{\texttt{HMCode}} (blue), or \texttt{Halofit} (orange). Black dashed lines mark the fiducial model.}
\label{fig:pla_wl}
\end{figure*}

We analyse this mock data set with a Markov Chain Monte Carlo method (MCMC), using the \texttt{MontePython}\footnote{\url{https://github.com/brinckmann/montepython_public}} suite \citep{Audren:2012wb,Brinckmann:2018cvx}, with a \Euclid lensing mock likelihood as presented in \citet{Sprenger:2018tdb}\footnote{With respect to \citet{Sprenger:2018tdb}, here we apply a $z$-independent cut-off at $k_{\rm max}=30\,h\,\mathrm{Mpc^{-1}}$, and we do not include the theoretical error.}. With this setting we sample the cosmological parameters listed in \autoref{tab:fiducialmcmc} using either \texttt{Halofit} (switching off \texttt{PKequal}) or \texttt{HMCode}. When using \texttt{HMCode}, the baryonic feedback parameters are fixed to the values fitting the \texttt{COSMIC EMU} dark-matter-only simulations \citep{coyote4}.

As done in \Cref{sec:FoMeffect}, we perform the analysis for $\ell_{\rm max}=1500$ and $5000$, in order to explore the impact of the nonlinear corrections on the results when including  high multipoles. Also in this case, we explore the difference with a scales cut in \Cref{sec:nlcut}. We complement the WL mock data set with TT, TE, EE, and lensing data from a mock \textit{Planck} likelihood, thus reproducing the sensitivity of the full mission. Note that we do not use real \textit{Planck} data to avoid a mismatch between the fiducial \Euclid cosmology and the actual best fitted \textit{Planck} values.

This approach enables us to determine the "bias" ($B$) on cosmological parameters with respect to our fiducial cosmology, i. e. the offset of the mean of the estimated posterior distribution from the true value, which in turn allows us to quantify the impact of using a different prescription for the nonlinear evolution of density perturbations. To assess the significance of a particular bias, it is useful to compare it to the expected statistical uncertainty. We therefore consider the (relative) bias on all the cosmological parameters given by
\begin{equation}\label{eq:biascomp}
B(\theta)=\frac{|\theta^\star-\theta^{\rm fid}|}{\sigma},
\end{equation}
where $\theta^\star$ is the mean value of the parameter found with the MCMC analysis, $\sigma$ is the $68\%$ uncertainty estimated from the chains, and $\theta^{\rm fid}$ is its fiducial value. We note that this estimate implicitly assumes that the posterior distribution can be approximated by a Gaussian, which is indeed valid for constraints based on the combination of \textit{Planck} and \Euclid data. 

The statistical uncertainties of \Euclid are determined by the survey design (see \Cref{sec:euspecs}). To draw reliable conclusions from these data, it is essential that systematic uncertainties are sufficiently small. Ideally biases should vanish, but generally it is too costly to achieve this. A reasonable compromise, however, is to adopt $B_{\rm T}\lesssim 0.1$ \citep[see e.g. Sect.~4.1 in][]{Massey13}, which is what we do here\footnote{This threshold could in principle be relaxed slightly if one wants to compromise for a lower variance. The investigation of such a trade off is however outside the scope of this paper.}. We note that the intrinsic variance of the mean estimated by the MCMC also contributes
to our estimate of $B(\theta)$. We quantified this contribution by computing the scatter of the mean value
by bootstrapping the chains. We find that it is always within $1\%$ of the final error on cosmological parameters.

The results are reported in \Cref{tab:pla_wl} and the posteriors are presented in \Cref{fig:pla_wl}. Although the optical depth $\tau$ is a free parameter in the model, it is effectively constrained by the \Planck measurements alone. We therefore do not report its value here. As expected, the biases are larger for \texttt{HMCode} compared to \texttt{Halofit} without \texttt{PKequal}. We also find that for \texttt{Halofit}, increasing the range from $\ell_{\rm max}=1500$ to $\ell_{\rm max}=5000$ does not increase the bias significantly, whereas
the bias strongly depends on the $\ell$-range for \texttt{HMCode}, both in amplitude and in sign. This can be explained by looking at \Cref{fig:pk_ratio}: at scales larger than a few $h\,\mathrm{Mpc^{-1}}$, \texttt{HMCode} systematically over-predicts the power with respect to \texttt{Halofit}.

We find that in the \texttt{Halofit} case, the biases in the  cosmological parameters approximately satisfy $B(\theta)\lesssim B_{\rm T}$ when $\ell_{\rm max}=1500$, while for \texttt{HMCode} the biases for almost all the parameters exceed this threshold. When setting $\ell_{\rm max}=5000$, the parameters estimated in the \texttt{HMCode} case are all biased significantly more than the acceptable threshold (except for $\omega_{\rm b,0}$), and now also the \texttt{Halofit} case exhibits biases larger than $B_{\rm T}$ for $h$, $\Omega_{\rm m,0}$ and $\sigma_8$.

In order to correct for the significant mismatch in the nonlinear prescriptions, \texttt{\texttt{HMCode}} increases $\omega_{{\rm c},0}$, $h$, 
$\ln(10^{10}A_{\rm s})$, and $w_a$, while at the same time the values for $n_{\rm s}$ and $w_0$ are decreased. This tweaking of parameters increases the amplitude of the linear matter power spectrum at scales $0.2\,h\,\mathrm{Mpc^{-1}}\lesssim k\lesssim2\,h\,\mathrm{Mpc^{-1}}$, where \texttt{\texttt{HMCode}} has a lack of power with respect to \texttt{\texttt{Halofit}+PKequal} (see \Cref{fig:pk_ratio}). As the scales around $0.2\,h\,\mathrm{Mpc^{-1}}$ are those that mainly contribute to the estimate of $\sigma_8$, this explains the large bias observed for this (derived) parameter, $B(\sigma_8)\sim 5$.

Overall, the $\Delta \chi^2 \lesssim 1$ indicates that replacing \texttt{Halofit+PKequal} with \texttt{Halofit}-only does not have a strong impact on the results, as it is well within the range of the statistical uncertainties\footnote{Note that as we do not introduce noise in our data vector, the $\chi^2$ for the fiducial model vanishes.}. On the other hand, using \texttt{HMCode} leads to a signficantly higher $\Delta\chi^2$, highlighting how the difference between the two nonlinear prescriptions cannot be fully compensated by modifying the background quantities and the linear growth.

It is worth to noting that for both \texttt{Halofit} and \texttt{HMCode} the parameters that are most significantly biased are $H_0$ and $\sigma_8$. These are the parameters that currently show tension between high- and low-redshift measurements \citep[e.g.][]{Riess:2019cxk,Hildebrandt:2018yau,SpurioMancini:2019rxy}. 
Our results imply that the \Euclid cosmic shear measurements have the statistical power to resolve this, but only if we can accurately model the nonlinear scales.

\begin{table*}[!htbp]
\centering
\caption{Mean values, marginalised $68\%$ errors, and biases in cosmological parameters. The values are obtained by fitting mock \textit{Planck} and \Euclid WL data to either \texttt{\texttt{HMCode}} without baryonic feedback or \texttt{Halofit} without \texttt{PKequal} nonlinear corrections. The last row shows $\Delta \chi^2$. By construction $\chi^2=0$, unless the configuration of the MCMC sampling does not match the one used to create the fiducial synthetic data set. The number of degrees of freedom in this case is 11 (the number of free parameters in the model; $\Omega_{\rm m,0}$ and $\sigma_8$ are derived parameters), which enables one to compare $\Delta\chi^2$ to the corresponding confidence interval.}
\begin{tabular}{lcccccccc} 
\hline
 \multicolumn{2}{c}{}& \multicolumn{3}{ c }{\texttt{Halofit}} && \multicolumn{3}{ c }{\texttt{\texttt{HMCode}}} \\
\cline{3-5}\cline{7-9}
$\theta$ & $\ell_{\rm max}$ & $\theta^\star$ &$\sigma$ &$B$ && $\theta^\star$ &$\sigma$ &$B$ \\
\hline
\hline
&$1500$ &$0.02244$ &$0.00011$ &$0.08$ &&$0.02240$ &$0.00012$ &$0.43$ \\ 
$\omega_{{\rm b},0}$ & & & & && & & \\ 
&$5000$ &$0.02243$ &$0.00012$ &$0.15$ &&$0.02246$ &$0.00012$ &$0.05$ \\ 
\hline 
&$1500$ &$0.12056$ &$0.00036$ &$0$ &&$0.12101$ &$0.00039$ &$1.16$ \\ 
$\omega_{{\rm c},0}$ & & & & && & & \\ 
&$5000$ &$0.12054$ &$0.00038$ &$0.053$ &&$0.12112$ &$0.00036$ &$1.57$ \\ 
\hline 
&$1500$ &$0.6689$ &$0.0069$ &$0.16$ &&$0.6702$ &$0.0096$ &$0.02$ \\ 
$h$ & & & & && & & \\ 
&$5000$ &$0.6683$ &$0.0048$ &$0.36$ &&$0.6899$ &$0.0066$ &$3.02$ \\
\hline 
&$1500$ &$3.0591$ &$0.0086$ &$0.09$ &&$3.0657$ &$0.0088$ &$0.84$ \\ 
$\ln(10^{10}A_\mathrm{s })$ & & & & && & & \\ 
&$5000$ &$3.0593$ &$0.0090$ &$0.10$ &&$3.0656$ &$0.0086$ &$0.85$ \\ 
\hline 
&$1500$ &$0.9602$ &$0.0025$ &$0.06$ &&$0.9615$ &$0.0027$ &$0.57$ \\ 
$n_\mathrm{s }$ & & & & && & & \\ 
&$5000$ &$0.9604$ &$0.0023$ &$0.18$ &&$0.9556$ &$0.0023$ &$1.90$ \\ 
\hline 
&$1500$ &$-0.888$ &$0.085$ &$0.14$ &&$-0.869$ &$0.099$ &$0.31$ \\ 
$w_{0}$ & & & & && & & \\ 
&$5000$ &$-0.888$ &$0.060$ &$0.21$ &&$-1.021$ &$0.064$ &$1.88$ \\ 
\hline 
&$1500$ &$0.07$ &$0.21$ &$0.14$ &&$-0.02$ &$0.25$ &$0.50$ \\ 
$w_{a}$ & & & & && & & \\ 
&$5000$ &$0.07$ &$0.16$ &$0.16$ &&$0.29$ &$0.16$ &$1.22$ \\ 
\hline 
&$1500$ &$0.3212$ &$0.0065$ &$0.18$ &&$0.3209$ &$0.0090$ &$0.10$ \\ 
$\Omega_\mathrm{{\rm m},0}$ & & & & && & & \\ 
&$5000$ &$0.32164$ &$0.0046$ &$0.36$ &&$0.3031$ &$0.0057$ &$2.96$ \\ 
\hline 
&$1500$ &$0.7852$ &$0.0058$ &$0.14$ &&$0.7938$ &$0.0071$ &$1.09$ \\ 
$\sigma_8$ & & & & && & & \\ 
&$5000$ &$0.7847$ &$0.0041$ &$0.30$ &&$0.8080$ &$0.0048$ &$4.62$ \\ 
\hline
\hline
             & $1500$ & \multicolumn{3}{ c }{$0.60$} &&\multicolumn{3}{ c }{$32.04$}\\
$\Delta\chi^2$ & & & & && & & \\
             & $5000$ & \multicolumn{3}{ c }{$1.06$} &&\multicolumn{3}{ c }{$62.34$}\\
\hline 
\end{tabular}

\label{tab:pla_wl}
\end{table*}

\section{Impact of baryons}\label{sec:bary}

Up to this point, we have limited our study to the impact of changing the recipe that is used to compute the nonlinear evolution of cold dark matter perturbations. On the small scales of interest, however, baryons collapse into the dark matter haloes to form stars, or are heated up, or even expelled into the intergalactic medium. These processes modify the matter distribution, and it is therefore important to account for baryonic physics when computing the matter power spectrum $P_{\delta\delta}(k,z)$ \citep[e.g.][]{vanDaalen:2011xb,casariniBar2,2018MNRAS.478.1305C,2020MNRAS.492.2285D}. This can be done by multiplying the nonlinear power spectrum, $P_{\delta\delta}(k,z)$ -- computed using one of the nonlinear prescriptions discussed above -- with ${\cal B}(k,z)$, a `baryon correction model' (BCM) that captures the baryonic effects \citep[e.g.][]{Semboloni11}, so that
\begin{equation}
P_{\rm c+b}(k,z) = P_{\delta\delta}(k,z) \mathcal B(k,z)\,,
\label{eq: bcmeq}
\end{equation}
where $P_{\rm c+b}(k,z)$ is the corrected power spectrum. The
function ${\cal B}(k,z)$ can be estimated by fitting \Cref{eq: bcmeq} to power spectra obtained from hydrodynamical simulations that include baryons.

The challenge is that baryonic effects cannot (yet) be incorporated into cosmological simulations from first principles. The different implementations that have been used, not surprisingly, lead to a variety of possible BCM prescriptions.  Here we consider three recent proposals. 

The first one presented in \citet[][HD15 hereafter]{Harnois-Deraps:2014sva}, is based on three scenarios of the  OverWhelmingly Large hydrodynamical simulations \citep{Schaye2010}. These were used to calibrate the power spectra for $z<1.5$. It is able to reproduce the simulated results 
with an accuracy better than 2\% for scales $k<1\,h\,{\rm Mpc}^{-1}$. The functional form of ${\cal B}(k,z)$ is given by
\begin{multline}
{\cal{B}}(k,z) = 1 - A_{\rm HD15}(z)\, \exp{\left \{ [B_{\rm HD15}(z)\, x(k) - C_{\rm HD15}(z)]^3 \right \}} \nonumber\\
+ D_{\rm HD15}(z)\, x(k)\,\exp\left[E_{\rm HD15}(z)\, x(k)\right]\,,
\label{eq:hd15}
\end{multline}
with $x(k) \equiv \log_{10}{(k/[h\,\mathrm{Mpc^{-1}}])}$ and $X_{\rm HD15}(z)$ are polynomial functions of redshift given in 
\cite{Harnois-Deraps:2014sva}.

As a second model, we consider the results obtained by \citet[][ST15 hereafter]{Schneider:2015wta}, who accounted for the effects of baryons following a different approach. They start from a suite of DM only N\,-\,body simulations, and modify the density field in such a way that it mimics the effects of a particular feedback recipe. They achieve this by explicitly modelling the main constituents of the halos, which are dark matter, hot gas in hydrostatic equilibrium, ejected gas and stars. The model parameters are set to resemble SZ and X\,-\,ray observations. The resulting modifications to the power spectrum are shown to be well reproduced by defining ${\cal B}(k,z)$ as
\begin{equation}
{\cal{B}}(k,z) = \frac{1 + (k/k_{\rm s})^2}{\left[ 1 + k/k_{\rm g}(z)\right]^3}\mathcal{G}(z) + \left[1 + (k/k_{\rm s})^2\right] \left[1 - {\cal{G}}(z)\right],
\label{eq: st15}
\end{equation}
with $k_{\rm g}(z)$ and ${\cal{G}}(z)$ auxiliary functions provided in
\cite{Schneider:2015wta}. We set the model parameters to the following fiducial values
\begin{equation}
\{k_{\rm s}, \log{M_{\rm c}}, z_{\rm b}, \eta_{\rm b}\} = \{67 \,h\, \mathrm{Mpc}^{-1}, 13.8, 2.3, 0.17\} \ . 
\end{equation}
Note that these are different from those of \citet{Schneider:2015wta}, since we have updated them to the best fitting values obtained using the more recent Horizon-AGN simulations \citep[as done in][]{Chisari:2018prw}. 

\citet[][Ch18 hereafter]{Chisari:2018prw} found that the ST15 model performs well at low redshift, but its accuracy degrades for larger $z$. Fitting the Horizon-AGN simulations, they therefore proposed the third model we will consider here, with
\begin{equation}
{\cal{B}}(k,z) = \frac{[1 + k/k_{\rm s}(z)]^2}{[1 + k/k_{\rm s}(z)]^{1.39}},
\label{eq: ch18}
\end{equation}
where $k_{\rm s}$ is no longer a constant, but a function of $z$ instead. The detailed form is given in \citet{Chisari:2018prw}.

We can now quantify the impact of the choice of BCM on the FoM by comparing it to the results for the dark-matter-only forecasts. To this end, we use \texttt{Halofit} as our benchmark model to compute $P_{\delta\delta}(k,z)$, which is consistent with what is done in the quoted papers. Our results are presented in \Cref{tab:FoMbar}.

The more recent Ch18 model yields the smallest change in the FoM, but the differences are never larger than $\sim 15\%$, even in the scenario with the largest change, i.e.\ the HD15 model with $\ell_{\rm max} = 5000$. This is the consequence of two opposite effects that partially cancel.  

On the one hand, at $k \sim 3$ -- $13\, h\,{\rm Mpc}^{-1}$, gas ejection due to AGN feedback suppresses the power spectrum, while for larger $k$ the effect of stars is to increase it. These very small scales $(k > 15\, h\, {\rm Mpc}^{-1})$ are, however, weighted down by the lensing kernel so that the overall effect is to reduce the signal in the $C^{\epsilon\epsilon}(\ell)$, which tends to reduce the FoM. However, reducing $C_{\ell}^{\epsilon\epsilon}$ also decreases the Gaussian covariance that is used to estimate the uncertainty in the WL signal, as the baryonic effects also change the power spectra used to compute the covariance matrix. As a result, the inverse covariance boosts the FoM. By moving the FoM in opposite directions, these two effects roughly compensate each other, so that the choice of the BCM model impacts the constraining power of the WL signal on dark energy parameters only marginally. Qualitatively similar results are obtained when we compare the marginalised uncertainties for individual parameters.  In general, we find that the parameter bounds are less affected than the FoM, in accordance with the argument given above.

\begin{table}[!h]
\centering
\caption{FoM for $w_0$ and $w_a$ parameters estimated using \Euclid specifications for weak lensing and the three BCM prescriptions compared to the \texttt{Halofit} forecast with no baryons.}
\begin{tabular}{lcccc} 
 \hline 
$\ell_{\rm max}$  & \texttt{Halofit} & HD15 & ST15 & Ch18  \\
\hline 
1500 & $23$  & $22$ & $21$ & $22$ \\
5000 & $44$ & $37$ & $41$ & $41$ \\
\hline 
\end{tabular}
\label{tab:FoMbar}
\end{table}

Although the detailed implementation of the BCM does not affect the constraining power much, it is nonetheless necessary to investigate whether an incorrect choice biases the cosmological parameter estimates. To this end, we adopt a procedure similar to that of \Cref{sec:mcmcbias}: we generate three simulated \Euclid WL data sets assuming that the BCM describing the true effect of baryons is CH18, ST15 or HD15, respectively. We then analyze these data sets, together with simulated \textit{Planck} CMB data, with theoretical predictions that neglect baryonic effects.

\begin{figure*}[!htbp]
\centering
\begin{tabular}{cc}
\includegraphics[width=0.48\textwidth]{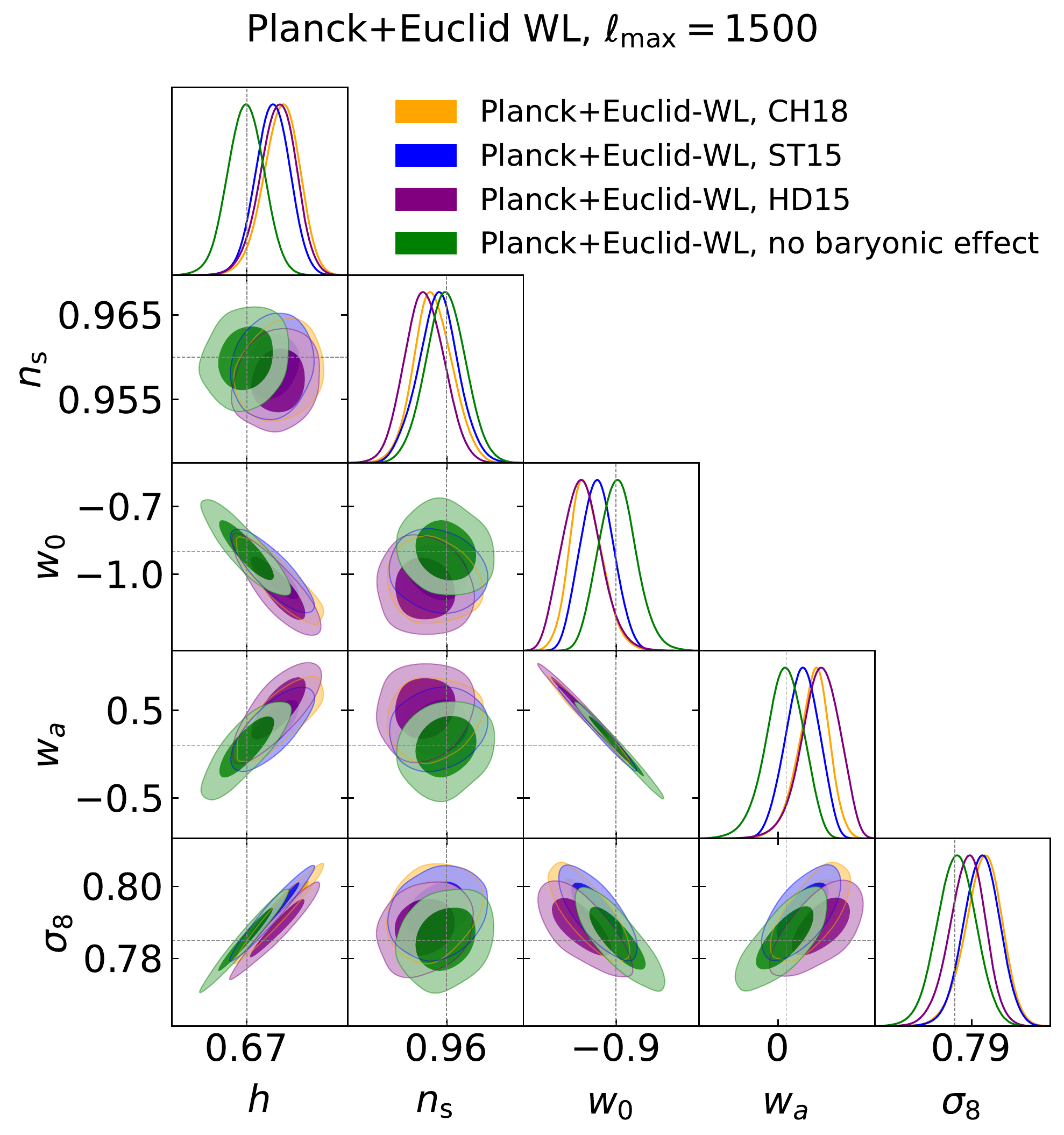}&
\includegraphics[width=0.48\textwidth]{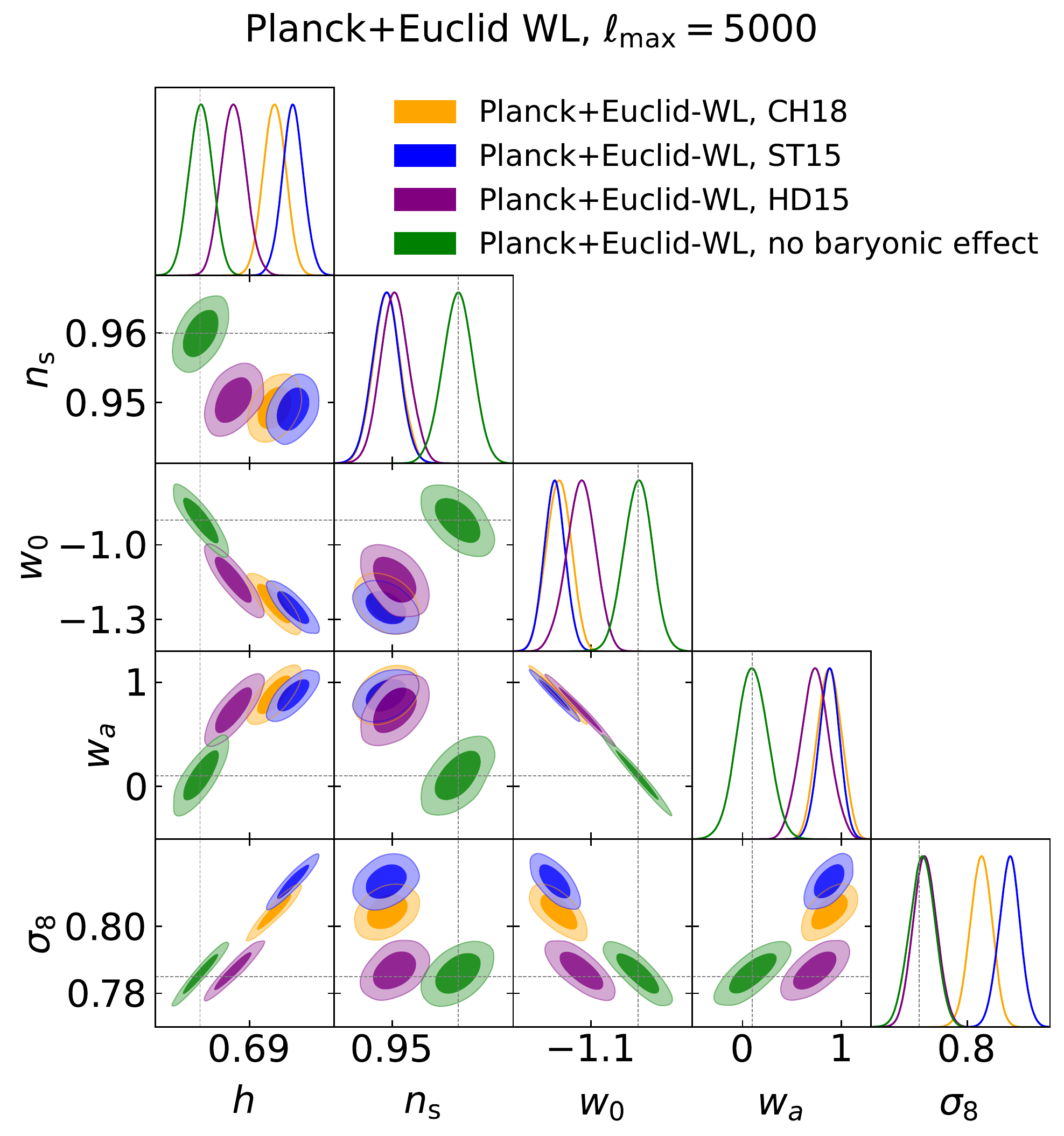}\\
\end{tabular}
\caption{One dimensional posterior distributions, and $68\%$ and $95\%$ marginalised joint two-parameter contours for $w_0$ and $w_a$, and the parameters $h$, $n_{\rm s}$ 
and $\sigma_8$ from the MCMC analysis. The results are obtained by neglecting baryon effects when fitting mock data sets created without baryonic effects (green) and with baryonic effects (orange for CH18, blue for ST15 and purple for Hd15).
The left panel refers to the $\ell_{\rm max}=1500$ case, while the right panel goes deeper into the nonlinear regime, with $\ell_{\rm max}=5000$.}
\label{fig:baryon_MCMC}
\end{figure*}

As done in the previous sections, we perform the MCMC analysis for both $\ell_{\rm max}$ values, $1500$ (pessimitic) and $5000$ (optimistic), which allows us to assess the relevance of the BCM for high multipoles. Our results are reported in \Cref{tab:pla_wl_baryons} and shown in \Cref{fig:baryon_MCMC}. For $\ell_{\rm max} = 1500$ we notice that the bias exceeds the threshold $B_{\rm T}$ for all parameters except for $\omega_{\rm b,0}$ and $\omega_{\rm c,0}$ in the CH18, while in ST15 also the bias on $\ln(10^{10}A_{\rm s})$ is lower than $B_{\rm T}$. HD15 instead only has this latter parameter within our acceptable threshold, a result confirmed by the highest $\Delta\chi^2$ among the three cases. The most significantly biased parameters in all three analysis are $h$, $w_0$ and $w_a$. Overall however, all three cases produce very similar results, as it can be seen in the left panel of \Cref{fig:baryon_MCMC}.

For $\ell_{\rm max} = 5000$, we find that the biases are very large when BCM effects are neglected; $B(\theta)>B_{\rm T}$ for all parameters, with $B\gtrsim5$ for the dark energy parameters $w_0$ and $w_a$, $B\gtrsim3$ for $h$,  
and $B\gtrsim4.5$ for $n_s$. As expected, the biases in the power spectrum amplitude, the baryon density and the cold dark matter density are the less significant, because these are all well constrained by the {\it Planck} measurements.

Ignoring BCM effects could lead to a false detection of a time-varying dark energy equation of state. Moreover, with the current tension between $H_0$ measurements between CMB and late-time probes, an unbiased measurement of $H_0$ will also be crucial.
Our results confirm earlier work \citep[e.g.][]{Semboloni11} that
correctly modelling the impact of baryonic feedback on the power spectrum is essential for the analysis of \Euclid data.

\begin{table*}[!htbp]
\centering
\caption{Mean values, marginalised $68\%$ errors, and bias. The values are obtained by fitting mock \textit{Planck} and \Euclid cosmic shear data with \texttt{Halofit} without baryonic corrections, to nonlinear corrections with either CH18, ST15 or HD15 methods nonlinear corrections. The number of degrees of freedom in this case is 11 (the number of free parameters), which enables one to compare $\Delta\chi^2$ to the corresponding confidence interval.}
\begin{tabular}{lcccccccccccc} 
\hline
 \multicolumn{2}{c}{} & \multicolumn{3}{ c }{\texttt{CH18}} && \multicolumn{3}{ c }{\texttt{ST15}}  && \multicolumn{3}{ c }{\texttt{HD15}} \\
 \cline{3-5}\cline{7-9}\cline{11-13}
$\theta$ & $\ell_{\rm max}$ & $\theta^\star$ &$\sigma$ &$B$ && $\theta^\star$ &$\sigma$ &$B$ && $\theta^\star$ & $\sigma$ & $B$ \\ 
\hline 
\hline 
$\omega_{{\rm b},0}$ & $1500$ & $0.2246$ & $0.00011$ & $0.09$ && $0.2245$ & $0.00012$ & $0.00$ && $0.02249$ & $0.00012$ & $0.34$ \\ 
             & $5000$ & $0.02251$ & $0.00012$ & $0.51$ && $0.02252$ & $0.00012$ & $0.64$ && $0.02252$ & $0.00012$ & $0.56$ \\ 
\hline
$\omega_{{\rm c},0}$ & $1500$ & $0.12061$ & $0.00035$ & $0.15$ && $0.12059$ & $0.00036$ & $0.08$ && $0.12044$ & $0.00040$ & $0.30$ \\  
             & $5000$ & $0.12109$ & $0.00036$ & $1.49$ && $0.12110$ & $0.00037$ & $1.46$ && $0.12083$ & $0.00036$ & $0.76$ \\ 
\hline             
$h$  & $1500$ & $0.6833$ & $0.0069$ & $1.92$ && $0.6799$ & $0.0065$ & $1.52$ && $0.6819$ & $0.0069$ & $1.73$ \\  
             & $5000$ & $0.6990$ & $0.0041$ & $7.15$ && $0.7069$ & $0.0041$ & $9.11$ && $0.6835$ & $0.0045$ & $2.99$ \\ 
\hline
$\ln(10^{10}A_{\rm s})$ & $1500$ & $3.0571$ & $0.0087$ & $0.14$ && $3.0590$ & $0.0083$ & $0.07$ && $3.0578$ & $0.0084$ & $0.07$ \\ 
             & $5000$  & $3.0644$ & $0.0083$ & $0.73$ && $3.0679$ & $0.0087$ & $1.10$ && $3.0592$ & $0.0083$ & $0.10$ \\ 
\hline
$n_{\rm s}$ & $1500$ & $0.9583$ & $0.0024$ & $0.68$ && $0.9589$ & $0.0025$ & $0.42$ && $0.9572$ & $0.0025$ & $1.11$ \\  
             & $5000$ & $0.9489$ & $0.0020$ & $5.49$ && $0.9488$ & $0.0021$ & $5.44$ && $0.9503$ & $0.0021$ & $4.69$ \\ 
\hline
$w_{0}$ & $1500$ & $-1.040$ & $0.078$ & $1.79$ && $-0.990$ & $0.078$ & $1.16$ && $-1.063$ & $0.092$ & $1.77$ \\ 
             & $5000$ & $-1.220$ & $0.039$ & $8.22$ && $-1.240$ & $0.040$ & $8.43$ && $-1.142$ & $0.053$ & $4.55$ \\ 
\hline
$w_{a}$ & $1500$ & $0.43$ & $0.19$ & $1.68$ && $0.30$ & $0.20$ & $0.98$ && $0.51$ & $0.23$ & $1.78$ \\ 
             & $5000$ & $0.84$ & $0.09$ & $8.27$ && $0.84$ & $0.10$ & $7.70$ && $0.73$ & $0.13$ & $5.04$ \\ 
\hline 
\hline
$\Delta\chi^2$ & $1500$ & \multicolumn{3}{ c }{$6.35$} && \multicolumn{3}{ c  }{$3.94$}  && \multicolumn{3}{ c }{$11.54$}\\
             & $5000$ & \multicolumn{3}{ c }{$63.61$} && \multicolumn{3}{ c  }{$107.32$} && \multicolumn{3}{ c }{$45.05$}\\
\hline 
\end{tabular}
\label{tab:pla_wl_baryons}
\end{table*}

\section{Conclusions}
\label{sec:conclusions}

Forthcoming surveys of the large-scale structure will deliver data sets of exquisite quality that will allow us to pursue what is usually referred to as {\it precision cosmology}. To correctly interpret these measurements significant improvements in the underlying theoretical predictions are needed: we need to ensure that errors in the modelling of density fluctuations on nonlinear scales do not introduce biases in the inferred cosmological parameters that are larger than the expected statistical uncertainties. This is a particular concern for cosmic shear tomography, given that one has to integrate the matter power spectrum $P_{\delta\delta}(k,z)$ deep into the nonlinear regime, where baryonic physics complicates matters even further. Motivated by these concerns, we have investigated how different popular prescriptions that account for these effects influence both the accuracy and the precision with which \Euclid can infer cosmological parameters using cosmic shear alone.

We used Fisher matrix forecasts to quantify the impact of three different nonlinear recipes on the dark energy figure of merit (FoM). The recipes that we considered are the revised implementation of \texttt{Halofit}, \texttt{Halofit+PKequal}, and the \texttt{HMCode} 
prescription. These differ significantly from one another when the cosmological parameters are left free to take values other than the fiducial ones. As a consequence, the derivatives of $P_{\delta\delta}(k,z)$ that enter the determination of the Fisher matrix are changed, leading to quite discrepant FoM values. In particular, we find that the \texttt{Halofit} case provides the higher FoM because of the different role played by $w_a$ in the nonlinear corrections. Although we have explicitly considered the case of \Euclid, this result is generic for cosmic shear tomography analyses, although the size of this effect will depend on the details of the survey of interest. Hence our findings highlight the importance of choosing the most reliable nonlinear model, in order to compute \emph{realistic} estimates of the expected performance of a particular weak lensing survey.

While it is important to quantify the {\it precision}, i.e. 
how tight the constraints will be, it is perhaps even more important to establish the {\it accuracy} of the results: we need to be confident that an incorrect choice of theoretical ingredients does not introduce an unacceptably large bias, i.e.\ a deviation from the (unknown) true value. Whether or not the bias is too large, also depends on the precision with which that parameter can be measured. 
We therefore define $B = |\theta - \theta^{\rm fid}|/\sigma$, and adopt a theshold of $B<B_{\rm T}\approx 0.1$.

To study the accuracy with which cosmological parameters can be determined, we created mock data with a given prescription for nonlinearities and/or baryon physics, and  fitting these with a different model. This allowed us to address
the issues of the choice of the nonlinear recipe and the baryon correction model separately. To examine the impact of the recipe used to compute the power spectra on nonlinear scales, we created a mock data set using \texttt{Halofit+PKequal} that comprises \Euclid cosmic shear and \textit{\textit{Planck}} CMB data. We emphasize that the choice of \texttt{Halofit+PKequal} is arbitrary, as we do not know which of the nonlinear corrections better describes the true small scale evolution. We fitted these with  \texttt{Halofit} or \texttt{HMCode}. 
 
We find that  $B\lesssim B_{\rm T}$ if \texttt{Halofit} is used for $\ell_{\rm max}=1500$, while for $\ell_{\rm max}=5000$, some cosmological parameters, namely $h$, $\Omega_{\rm m,0}$ and $\sigma_8$, exceed the threshold. This is not surprising given the similarities of the two models when one only looks at $P_{\delta\delta}$ rather than at its derivatives. In contrast,
the use of \texttt{HMCode} leads to strong biases, with almost all parameters already biased by more than $B_{\rm T}$ if we restrict the analysis to $\ell_{\rm max}=1500$. Including the very nonlinear regime scales($\ell_{\rm max} = 5000$), we find that $B>1$ for all parameters, except for $\omega_{{\rm b},0}$, which is actually tightly constrained by {\it Planck}. In particular, the estimate of $w_0$ shifts towards its $\Lambda$CDM value even if the mock data were created using $\{w_0,\,w_a\} = \{-0.9,\,0.1\}$. What is even more interesting is that the most biased parameters are $h$ 
($B=3.02$) and $\sigma_8$ ($B=4.62$); the values of both of these are currently debated. The sensitivity of these parameters to the adopted prescription for the nonlinear power spectrum highlights the need for further improvements, which may already be needed to correctly interpret current data. 

It is also essential that the changes to the power spectrum caused by baryon physics have to be taken into account. For a fixed nonlinear recipe and baryon correction model prescription, severe biases are found when fitting the mock data with the right nonlinear correction, but not accounting for the presence of baryons, in line with earlier work \citep[e.g.][]{Semboloni11}. In the most constraining setting ($\ell_{\rm max}=5000$), for all the three cases we considered, we find significant biases for all the cosmological parameters, except for $\{\omega_{{\rm b},0}, \ln{(10^{10} A_{\rm s})}\}$, which are actually constrained by the \textit{\textit{Planck}} data rather than by cosmic shear. While this work was near to completion, \citet{Schneider:2019xpf} presented a similar analysis, but their method to account for baryons is very different from the one we have adopted here. They use instead a model for the {\it baryonification} of dark matter only simulations to determine the matter power spectrum \citep{Schneider:2018pfw}. Notwithstanding these differences, which make a straightforward comparison impossible, their conclusions are in agreement with what we found here. 

As a final remark, we remind the reader that our results refer to the case where cosmic shear is used as the only probe. This is, however, only part of the information that future surveys will provide. Indeed, the same data used to do cosmic shear tomography (WL) can and will be used to compute the photometric galaxy clustering (GCph) and to cross correlate the shear and density fields (XC).  As was shown in \citetalias{IST:paper1} and further investigated in \citet{Tutusaus:2020xmc}, it is the joint use of WL+GCph+XC that is needed to achieve $\sim \{1, 10\}\%$ errors on $\{w_0, w_a\}$, rather than any single probe by itself. It is therefore worth considering whether, and to which extent, the results we have obtained here change if all the three probes are considered. 

For instance, \citetalias{IST:paper1} limited the GCph and XC to smaller multipoles compared to what was used for WL, reducing
their sensitivity to the small scales corrections. This diminishes the impact of errors in the prediction of $P_{\delta\delta}(k,z)$ in the large $k$ regime, but at the expense of larger statistical uncertainties. This motivates extending our work to quantify the impact of modelling the small scale power spectra for the joint probes. Such a study, which is beyond the scope of our initial exploration, would provide guidance how to proceed in order to exploit the high-quality data that stage IV surveys will provide.

\appendix

\section{Comparison between scales and multipoles cuts}\label{sec:nlcut}

In the analyses performed in this paper, we considered two cases, with different cuts in multipoles used; the optimistic $\ell_{\rm max}=5000$ cut represents the situation in which data for all scales are included in the analysis, while the $\ell_{\rm max}=1500$ mimics the removal of nonlinear scales that could be performed in the data analysis in order to reduce the impact of small scales modeling.

However, our approach can be seen as a first approximation, as a constant multipole cut corresponds to different scales for different redshifts. In this subsection we investigate a less simplistic approach, implementing a $k_{\rm max}$ scale cut, rather than a multipole one, with the purpose of removing part of the nonlinear scales.

We assume $k_{\rm max}=0.25\ h$ Mpc$^{-1}$ and, using the Limber approximation considered to express the $C(\ell)$ in \Cref{eq:cells}, we convert this cut in a maximum multipole for each redshift bin:
\begin{equation}
 \ell_{\rm max}(z_i) = k_{\rm max}r(z_i)-\frac{1}{2}\, ,
\end{equation}
where $z_i$ is the mean redshift of each redshift bin.

\begin{figure*}[!htbp]
\centering
\begin{tabular}{cc}
\includegraphics[width=0.48\textwidth]{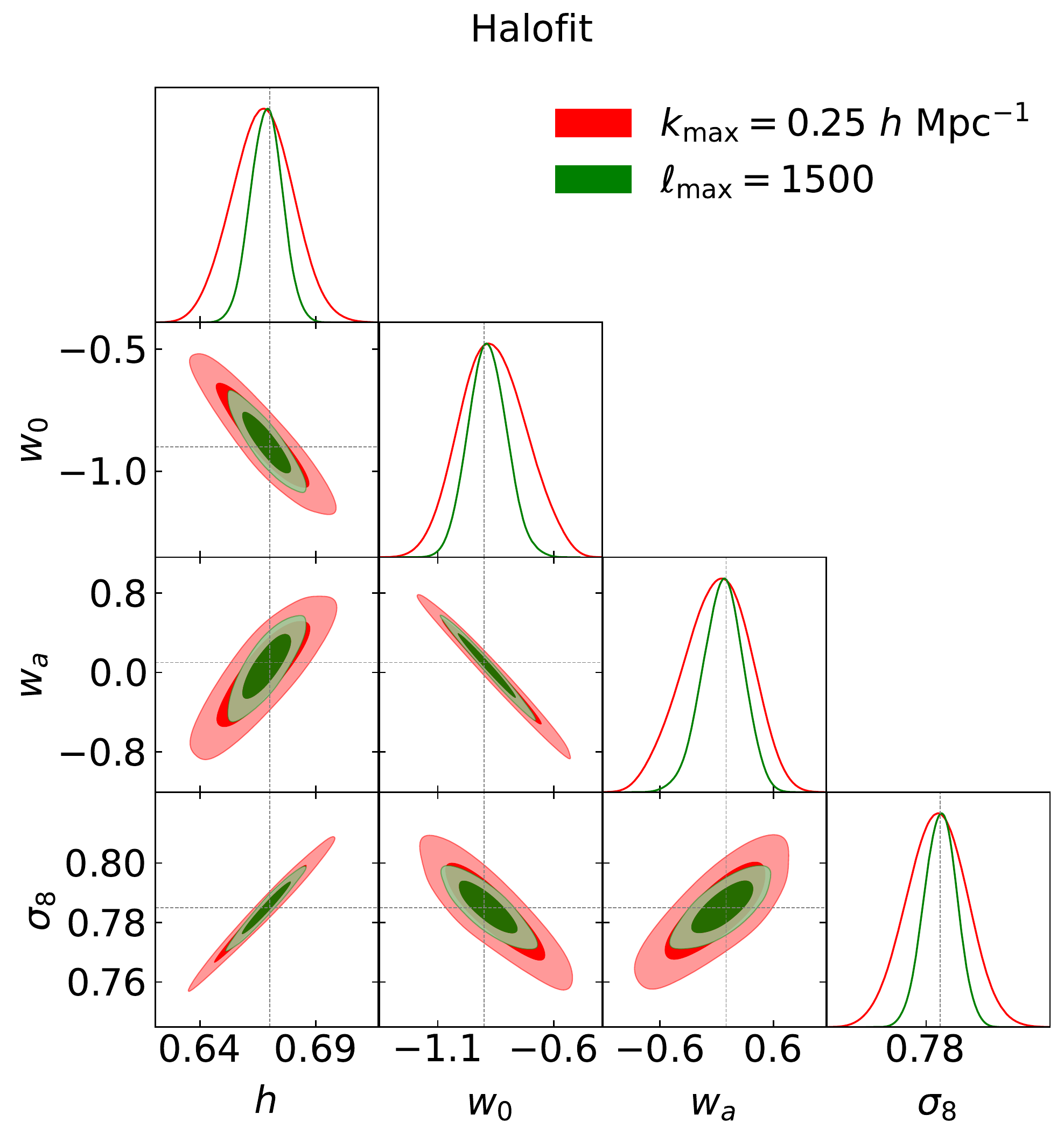}&
\includegraphics[width=0.49\textwidth]{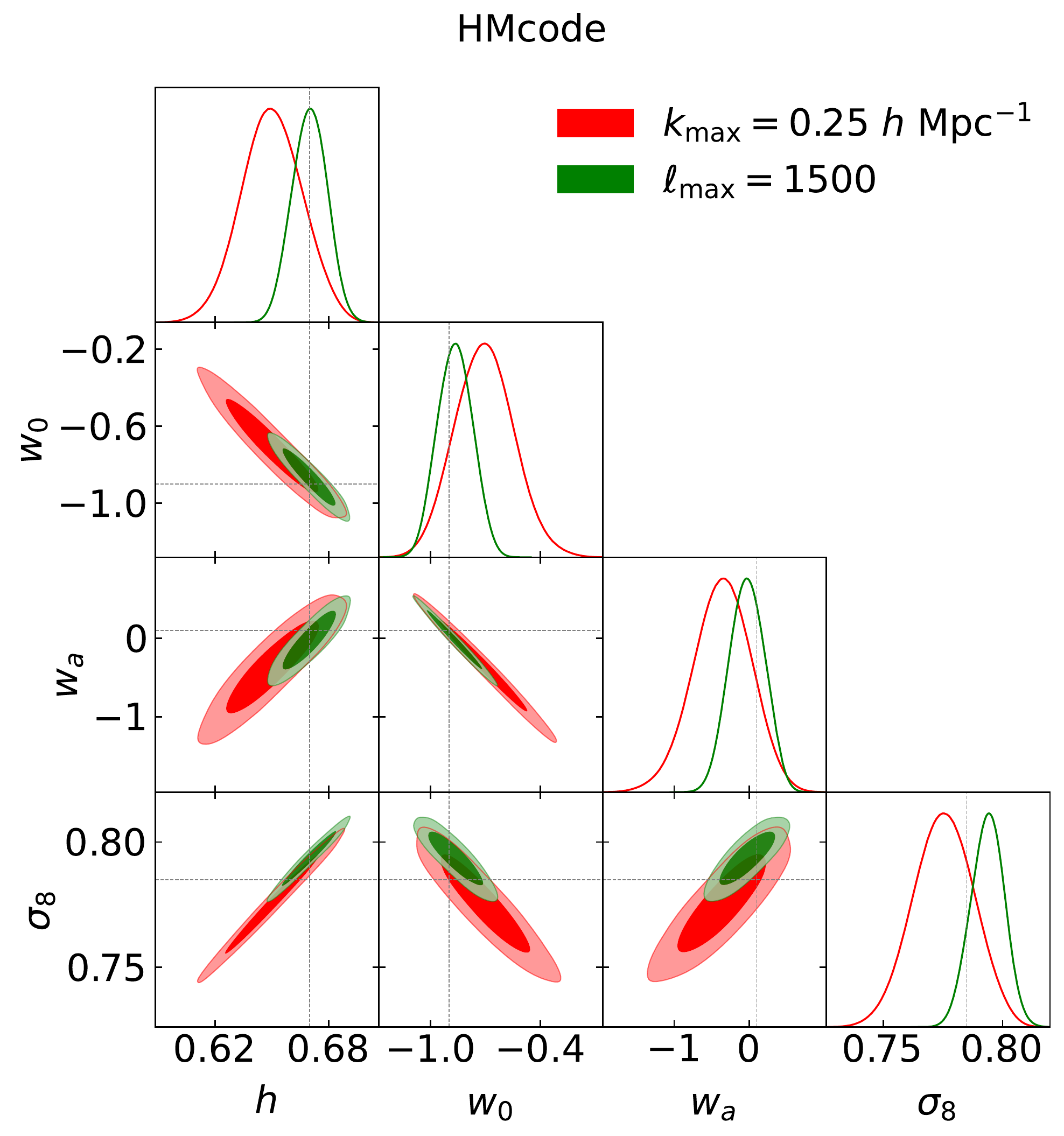}\\
\end{tabular}

\caption{One dimensional posterior distributions, and $68\%$ and $95\%$ confidence level marginalised contours for the dark energy parameters ($w_0$ and $w_a$) and the parameters $h$
and $\sigma_8$. Here we compare results obtained with a multipole cut at $\ell_{\rm max}=1500$ (green contours) with those related to the analysis with the scales cut $k_{\rm max}= 0.25\ h$ Mpc$^{-1}$ (red contours). The mock data of \Euclid cosmic shear assume \texttt{\texttt{Halofit}+PKequal} nonlinear corrections as reference, while the parameter estimation is performed with either \texttt{\texttt{HMCode}} (right panel), or \texttt{Halofit} (left panel). Black dashed lines mark the fiducial model.}
\label{fig:kcut}
\end{figure*}

We apply this approach to our MCMC analysis, and compare the results we obtain with those of the $\ell_{\rm max}=1500$ case. In \Cref{fig:kcut} we show the comparison in the results for the analysis performed using \texttt{Halofit} (left panel) and \texttt{HMCode} (right panel). In both cases we find that the $k_{\rm max}$ analysis leads to broader constraints with respect to the $\ell_{\rm max}$ case; this is due to the fact that the chosen $k_{\rm max}$ translates into a much more agressive cut in multipoles, specially at low redshift. 
Therefore, a significative amount of information is lost with respect to the $\ell_{\rm max}=1500$ case.

Concerning the bias found on cosmological parameters, $B(\theta)$ does not change significantly in the \texttt{Halofit analysis}, while for \texttt{HMCode} the biases are slightly enhanced with respect to the $\ell_{\rm max}=1500$ case, with $w_0,\ w_a,\ h$ and $\sigma_8$ now reaching $B\approx1$. This apparent increase however is mostly due to the loss of constraining power when the scale cut is implemented; as it can be seen in \Cref{fig:kcut}, the marginalized posterior distributions for the cosmological parameters now exhibit non-Gaussian features while \Cref{eq:biascomp} implicitly assumes Gaussian distributions. In \Cref{fig:kcut} it is shown how the scale cut case produces contours closer to the expected fiducial values, a result supported also by the change in $\Delta\chi^2$ moving from the $\ell_{\rm max}$ to the scale cut, which changes from $32$ to $14$ in the \texttt{HMCode} case.

We have performed the same analysis at the Fisher matrix level finding a still significant dependence of the results on the adopted nonlinear recipe. We indeed get FoM $= (2.93, 3.59, 1.59)$ for {\tt Halofit}, {\tt HMCode}, {\tt Halofit\,+\,PKequal}, respectively. While the severe decrease of the FoM is expected given that we are removing a large part of the data, it is somewhat surprising to still find such a variety of values. This is, however, a consequence of the integrated nature of the WL $C_{ij}(\ell)$. To understand what is going on, let us focus on the case $i = j = 5$ giving $\ell_{\rm max} \simeq 500$. Because of the photo\,-\,z broadening of the lensing kernel, the integral giving $C_{55}(\ell_{\rm max})$ gets contributions from the redshift range $(0.1, 2.6)$. Over this range, the argument $k_\ell(z)$ of the matter power spectrum $P_{\delta \delta}(k, z)$ feeding the integral is larger than $k_{\rm max}$ for $z < 0.84$ so that which nonlinear recipe is adopted still matters. Such an argument can be repeated for all the bins combinations and the multipoles thus explaining why the FoM is still dependent on the nonlinear recipe even with these very conservative scale cuts, a result in agreement with what was discussed in \citet{Taylor:2018snp}. Therefore, in order to remove completely the dependence on the nonlinear description from the analysis, different approaches are needed, e.g. using band powers rather than a $C_\ell$ analysis \citep{Joachimi:2020abi}. 

\section{MCMC results validation}\label{app:mcmcvalidation}

The results of this paper have been obtained using both Fisher matrix and MCMC codes.   The Fisher analysis relies on one of the codes used in \citetalias{IST:paper1}, which has passed through a careful validation procedure that included intensive comparisons between different Fisher matrix codes.  Our MCMC analysis relies on a public \texttt{MontePython} likelihood for \Euclid WL \citep{Brinckmann:2018cvx}, adapted to the \Euclid specifications of \citetalias{IST:paper1}. It has been further modified to include different models of baryonic feedback effects. This likelihood code was first used in \citet{Sprenger:2018tdb}. In contrast to the Fisher code, it has not been validated against other codes. 
In this Appendix, we therefore present a comparison between validated Fisher forecasts and the MCMC ones.  

In the \Euclid-only case, our analysis reveals some deviations that are attributed to the intrinsic limitation of any Fisher analysis, due to the non-Gaussianity of the posterior distribution, as well as some important parameter degeneracies.  Nevertheless, the impact on the forecasts becomes negligible when \Planck constraints are included in the analysis. In this case, we find that the forecasts on cosmological parameters obtained with the Fisher and MCMC methods agree very well. This therefore validates our MCMC approach against the Fisher codes used in \citetalias{IST:paper1}.

The \Euclid forecasts for WL in \citetalias{IST:paper1} are accompanied by a series of public Fisher matrices, corresponding to different setups and cosmologies.  In order to validate our \texttt{MontePython} likelihood and MCMC analysis, we have compared forecasts obtained for both the pessimistic and optimistic setups, and using the same cosmological parameters (in particular, $\Omega_{\rm b,0} $ and $\Omega_{\rm m,0}$ instead of $\omega_{\rm b}$ and $\omega_{\rm c}$).  We have performed these comparisons for \Euclid only, and in combination with {\it Planck}.  In the latter case, we used the mock \Planck likelihood available in \texttt{Montepython} that accurately reproduces  the \Planck limits on cosmological parameters. We construct a covariance matrix from the MCMC chains in the {\it Planck}-only case.  Its inverse provides a Fisher matrix that can be added to the validated \Euclid Fisher matrices.  We have checked that the {\it Planck}-only case constrains the standard cosmological parameters well, with close-to-Gaussian two-dimensional posterior distributions.  This is a good indication that one can safely use the {\it Planck} covariance matrix for the Fisher analysis.  In contrast to the standard cosmological parameters, most of the constraining power for the dark energy parameters $w_0$ and $w_a$ comes from \Euclid, not {\it Planck}.  As a consequence, the corresponding entries in the {\it Planck} Fisher matrix are not relevant and do not significantly impact the \Euclid+{\it Planck} forecasts for these parameters.

\begin{figure*}
\centering
\includegraphics[width=0.99\textwidth]{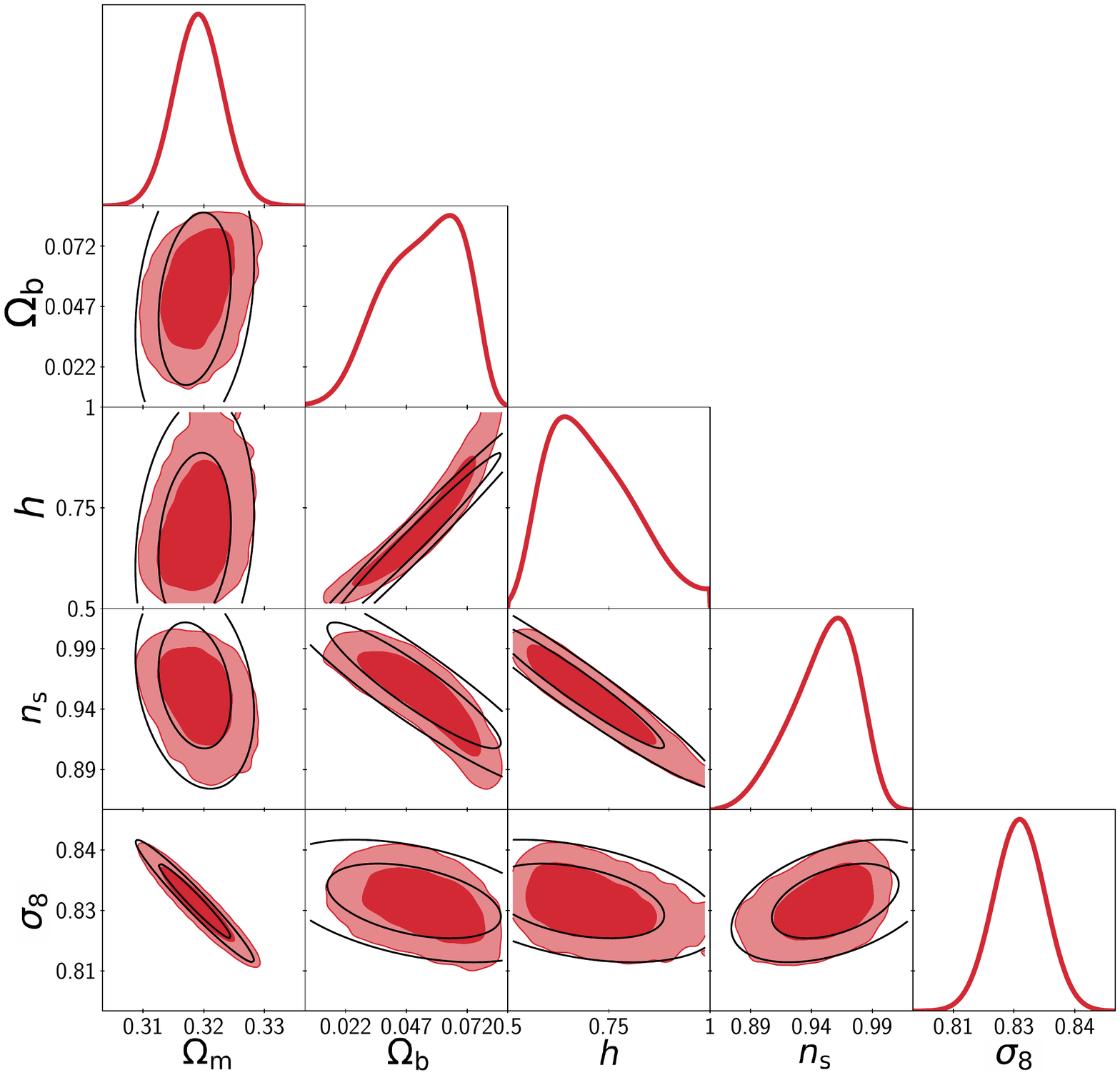}
\caption{Fisher matrix vs MCMC forecasts, for \Euclid WL only with $\ell_{\rm max} = 1500$.   In red, one dimensional marginalized posterior distributions, with two-dimensional $68\%$ and $95\%$ confidence level marginalised contours, for varying cosmological parameters and fixed DE and IA parameters.  Black curves represent the corresponding Fisher ellipses, from the validated Fisher matrix of \citetalias{IST:paper1}. }
\label{fig:fishervsMCMC1}
\end{figure*}

For \Euclid only, the marginalized two-dimensional posterior distributions and the Fisher contours obtained in the case $\ell_{\rm max}=1500$ for five varying cosmological parameters ($\Omega_{\rm m,0}$, $\Omega_{\rm b,0}$, $h$, $n_{\rm s}$ and $\sigma_8$), with DE and IA parameters fixed to their fiducial values, are shown in Fig.~\ref{fig:fishervsMCMC1}.  We find that the directions and widths of all the Fisher ellipses are well recovered by the MCMC approach.  However, we also find two significant differences.  First, the MCMC contours in the plane $(\Omega_{\rm b,0},h)$ display a \textit{banana} shape, leading to a more stringent constraint on $\Omega_{\rm b,0}$ compared to the Fisher method.  In turn, this affects two-dimensional contours between $\Omega_{\rm b,0}$ and other parameters.  Second, the marginalized posterior distribution for $h$ has a significant non-Gaussian shape, falling more rapidly at lower values than at higher values.  Consequently, some of the contours are less extended on one side compared to the Fisher ellipses. This also arises from the degeneracy between $h$ and $\Omega_{\rm b,0}$, with these parameters being poorly constrained with \Euclid WL only.  These features cannot be recovered by the Fisher analysis that relies on the assumption that the posterior distribution is Gaussian, which is not the case for strong variations of $h$ and $\Omega_{\rm b,0}$. Nonetheless, these results show that both the MCMC and Fisher methods provide consistent results, even if the agreement between them is limited by the intrinsic limitation of the Fisher analysis.  Adding dark energy parameters in the analysis further degrades the agreement between the Fisher and MCMC forecasts, for similar reasons.  Here, we have only shown the case in which $w_0$ and $w_{\rm a}$ remain fixed, for a better illustration of the effect of parameter degeneracies on the two method comparison. 

\begin{figure*}
\centering
\includegraphics[width=0.99\textwidth]{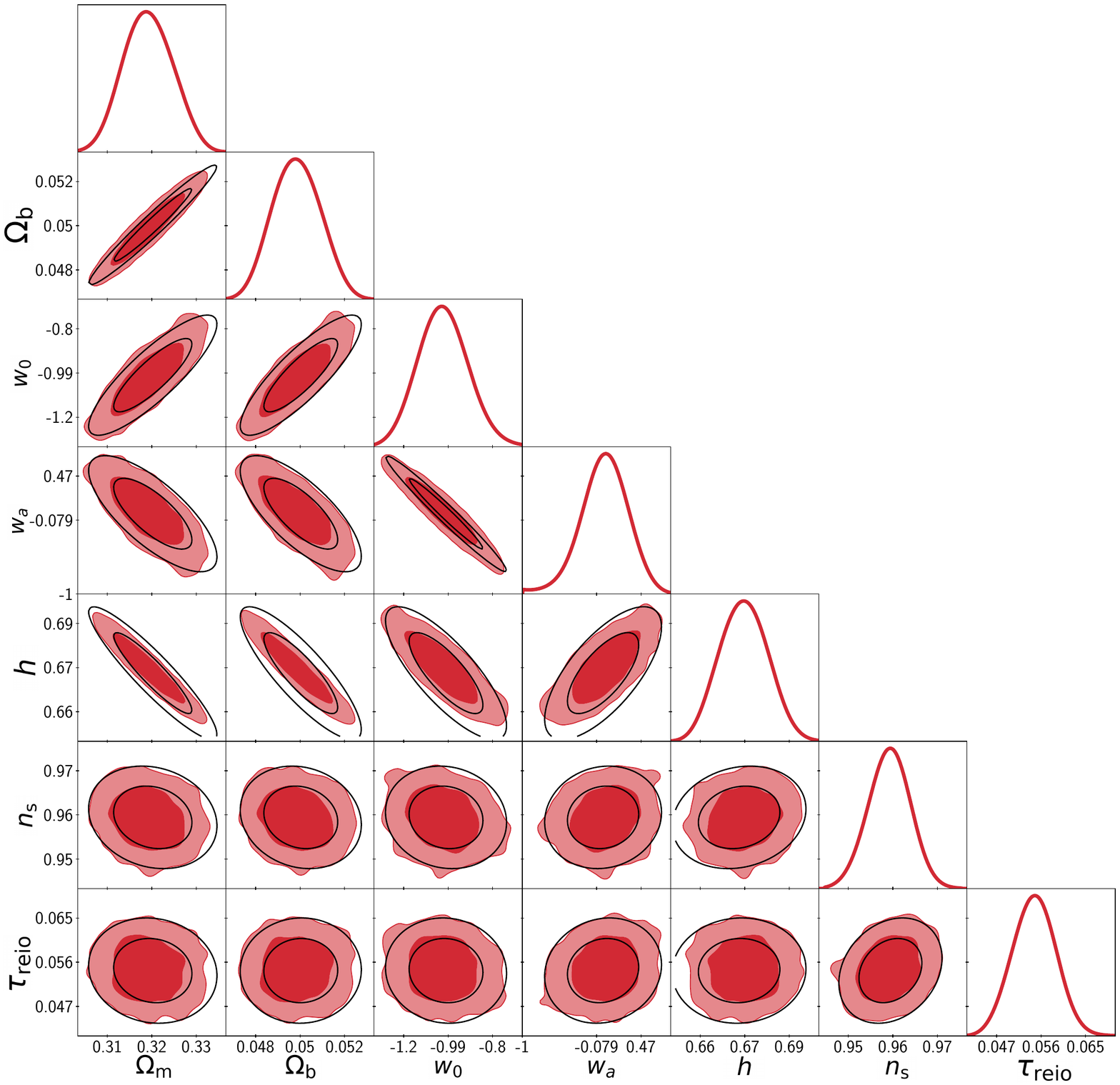}
\caption{Fisher matrix vs MCMC forecasts, for {\it Planck}+\Euclid WL, with $\ell_{\rm max} = 1500$.  In red, one dimensional marginalized posterior distributions, with two dimensional $68\%$ and $95\%$ confidence level marginalised contours, for varying cosmological parameters including $w_0$ and $w_{\rm a}$ and IA parameters.  Black curves represent the corresponding Fisher ellipses, from the validated Fisher matrix of \citetalias{IST:paper1} combined with a {\it Planck}-only Fisher matrix. }
\label{fig:fishervsMCMC2}

\end{figure*}

For {\it Planck}+\Euclid WL, we find a very good agreement between Fisher and MCMC forecasts, even when the DE and IA parameters are varying.  This is shown in Fig.~\ref{fig:fishervsMCMC2} representing the MCMC marginalized two-dimensional posterior distributions, which fit well to all the Fisher ellipses.  We find that the differences in the forecasts are less than $10\%$ for all the cosmological parameters.  This therefore validates the approach we have used throughout the paper.  It also illustrates the importance of adding CMB data to break parameter degeneracies.  Compared to the \Euclid WL only case, the differences between the Fisher and MCMC methods are suppressed.  The few remaining differences can be either due to a limited convergence of the MCMC chains, to uncertainties induced by the binning method, or to a slightly non-Gaussian likelihood function for {\it Planck}-only that induces small differences between MCMC posteriors and the corresponding {\it Planck} Fisher ellipses extracted from the covariance matrix of the MCMC chains.   Similar results have been obtained for the optimistic setup with $\ell_{\rm max}=5000$.

\begin{acknowledgements}
DFM thanks the Research Council of Norway for their support, and the resources provided by
UNINETT Sigma2 -- the National Infrastructure for High Performance Computing and 
Data Storage in Norway. MM has received the support of a fellowship from "la Caixa" Foundation (ID 100010434), with fellowship code LCF/BQ/PI19/11690015, and the support of the Spanish Agencia Estatal de Investigacion through the grant “IFT Centro de Excelencia Severo Ochoa SEV-2016-0597”. This paper is based upon work from the COST action CA15117 (CANTATA), supported by COST (European Cooperation in Science and Technology). 
MA acknowledges the computing support of INFN-CNAF.
Stefano Camera acknowledges support from the Italian Ministry of Education, University and Research (\textsc{miur}) through Rita Levi Montalcini project `\textsc{prometheus} -- Probing and Relating Observables with Multi-wavelength Experiments To Help Enlightening the Universe's Structure', and the `Departments of Excellence 2018-2022' Grant awarded by \textsc{miur} (L.\ 232/2016).  The work of SC is supported by the Belgian Fund for Research FNRS-F.R.S. IT acknowledges support from the Spanish Ministry of Science, Innovation and Universities through grant ESP2017-89838-C3-1-R. IT and TK acknowledge funding from the H2020 programme of the European Commission through grant 776247. AP is a UK Research and Innovation Future Leaders Fellow, grant MR/S016066/1. \AckEC
\end{acknowledgements}

\bibliographystyle{aa}
\bibliography{references}

\end{document}